\documentclass[twocolumn]{aastex631}

\usepackage{textgreek, gensymb, needspace}
\usepackage[]{fixltx2e}
\usepackage[bottom]{footmisc}
\received{March 1, 2023}
\revised{April 28, 2023}
\accepted{May 16, 2023}

\submitjournal{AJ}

\shorttitle{Active low-mass M dwarfs}
\shortauthors{Pass et al.}

\begin{document}

\title{Active Stars in the Spectroscopic Survey of Mid-to-Late M Dwarfs Within 15pc}

\author[0000-0002-1533-9029]{Emily K. Pass}
\affiliation{Center for Astrophysics $\vert$ Harvard \& Smithsonian, 60 Garden Street, Cambridge, MA 02138, USA}

\author[0000-0001-6031-9513]{Jennifer G. Winters}
\affiliation{Center for Astrophysics $\vert$ Harvard \& Smithsonian, 60 Garden Street, Cambridge, MA 02138, USA}
\affiliation{Thompson Physics Lab, Williams College, 880 Main Street, Williamstown, MA 01267, USA}

\author[0000-0002-9003-484X]{David Charbonneau}
\affiliation{Center for Astrophysics $\vert$ Harvard \& Smithsonian, 60 Garden Street, Cambridge, MA 02138, USA}

\author{Jonathan M. Irwin}
\affiliation{Center for Astrophysics $\vert$ Harvard \& Smithsonian, 60 Garden Street, Cambridge, MA 02138, USA}
\affiliation{Insitute of Astronomy, University of Cambridge, Madingley Road, Cambridge CB3 0HA, UK}

\author[0000-0001-8726-3134]{Amber A. Medina}
\affiliation{Center for Astrophysics $\vert$ Harvard \& Smithsonian, 60 Garden Street, Cambridge, MA 02138, USA}
\affiliation{Department of Astronomy, The University of Texas at Austin, 2511 Speedway, Austin, TX 78712, USA}



\begin{abstract}
We present results from the volume-complete spectroscopic survey of 0.1--0.3M$_\odot$ M dwarfs within 15pc. This work discusses the active sample without close binary companions, providing a comprehensive picture of these 123 stars with H\textalpha\ emission stronger than -1\AA. Our analysis includes rotation periods (including 31 new measurements), H\textalpha\ equivalent widths, rotational broadening, inclinations, and radial velocities, determined using high-resolution, multi-epoch spectroscopic data from the TRES and CHIRON spectrographs supplemented by photometry from TESS and MEarth. Using this volume-complete sample, we establish that the majority of active, low-mass M dwarfs are very rapid rotators: specifically, 74$\pm$4\% have rotation periods shorter than 2 days, while 19$\pm$4\% have intermediate rotation periods of 2--20 days, and the remaining 8$\pm$3\% have periods longer than 20 days. Among the latter group, we identify a population of stars with very high H\textalpha\ emission, which we suggest is indicative of dramatic spindown as these stars transition from the rapidly to slowly rotating modes. We are unable to determine rotation periods for six stars and suggest that some of the stars without measured rotation periods may be viewed pole-on, as such stars are absent from the distribution of inclinations we measure; this lack notwithstanding, we recover the expected isotropic distribution of spin axes. Our spectroscopic and photometric data sets also allow us to investigate activity-induced radial-velocity variability, which we show can be estimated as the product of rotational broadening and the photometric amplitude of spot modulation. 
\end{abstract}


\section{Introduction} \label{sec:intro}

M dwarfs are observed to have a bimodal magnetic activity distribution of either saturated or unsaturated behavior. Young, active M dwarfs are in the saturated regime, where rotation rate is uncorrelated with a variety of activity proxies such as X-ray emission \citep{Wright2011, Wright2018}, H\textalpha\ luminosity \citep{Newton2017}, UV emission \citep{France2018}, and flares \citep{Medina2020, Medina2022}. M dwarfs that rotate more slowly are in the unsaturated regime, where activity lessens with lengthening rotation periods. These works estimate the Rossby number of the transition to be somewhere between 0.1 and 0.5, corresponding to rotation periods of 10--50 days for a 0.2M$_\odot$ M dwarf \citep{Wright2018}. Importantly, mid-to-late M dwarfs remain in the saturated regime for gigayears, with an estimated average epoch of spindown of 2.4 $\pm$ 0.3 Gyr \citep[][although there can be a large variability in this epoch; e.g., \citealt{Pass2022}]{Medina2022}. As low-mass M dwarfs remain active for such an extended period, these active stars are a significant stellar demographic, with particular relevance to studies of exoplanets and planetary habitability \citep[e.g.,][]{Lammer2007, Tilley2019}; that is, the M dwarfs that are old and inactive today were once these active stars, and their extant planets formed and evolved during this lengthy phase of activity.

In addition, while planets can be discovered and their masses determined through the radial-velocity (RV) perturbations they induce on their host star, stellar activity also generates similar signals. The quantitative relationship between starspots and RV variation was first explored in \cite{Saar1998} and has been investigated in greater detail in the years since \citep[][]{Lanza2011, Aigrain2012, Boisse2012, Haywood2014, Oshagh2017, Hojjatpanah2020, Baroch2020, Jeffers2022}. For young, active M dwarfs, the amplitude of the activity-induced RV jitter can be tens to hundreds to even thousands of meters per second \citep[e.g.,][]{Tal-Or2018}. Such a star is considered to be ``RV loud.'' Stellar activity therefore complicates searches for exoplanets.

This work is part of a series of papers presenting the results of the volume-complete spectroscopic survey of 0.10--0.30M$_\odot$ M dwarfs within \hbox{15pc}, defined in \citet{Winters2021}. This mass range roughly corresponds to spectral types M4V--M7V. The sample totals 413 stars and excludes M dwarfs that are close companions to more massive primaries. The single, inactive subsample was discussed in \citet{Pass2023}, where we used these stars to place constraints on the occurrence rate of giant planets around low-mass M dwarfs. For the subset of the single-star sample that was observed during the primary mission of the Transiting Exoplanet Survey Satellite (TESS), \citet{Medina2020, Medina2022} published flare rates, rotation periods, and spectroscopic activity indicators. The time-dependent H\textalpha\ variability of thirteen of these stars was also studied in \citet{Medina2022a}. In this work, we present results for the single, active subsample, which we define as the 123 stars without a binary companion within 4" and with H\textalpha\ emission stronger than a median equivalent width of \hbox{-1\AA}. This \hbox{-1\AA} threshold has been used to distinguish between active and inactive M dwarfs in previous work such as \citet{Newton2017}.

In Section~\ref{sec:reduction}, we describe our spectroscopic analysis. In Section~\ref{sec:rot}, we collate complementary photometric data and present rotation periods for the majority of our sample. In Section~\ref{sec:i}, we combine the spectroscopic and photometric data sets to measure inclinations. In Section~\ref{sec:rv}, we present our multi-epoch RV measurements and discuss activity-induced RV variability. We conclude in Section~\ref{sec:conclusion}.

\begin{figure*}[h!]
    \centering
    \makebox[\textwidth][c]{\includegraphics[width=1.\textwidth]{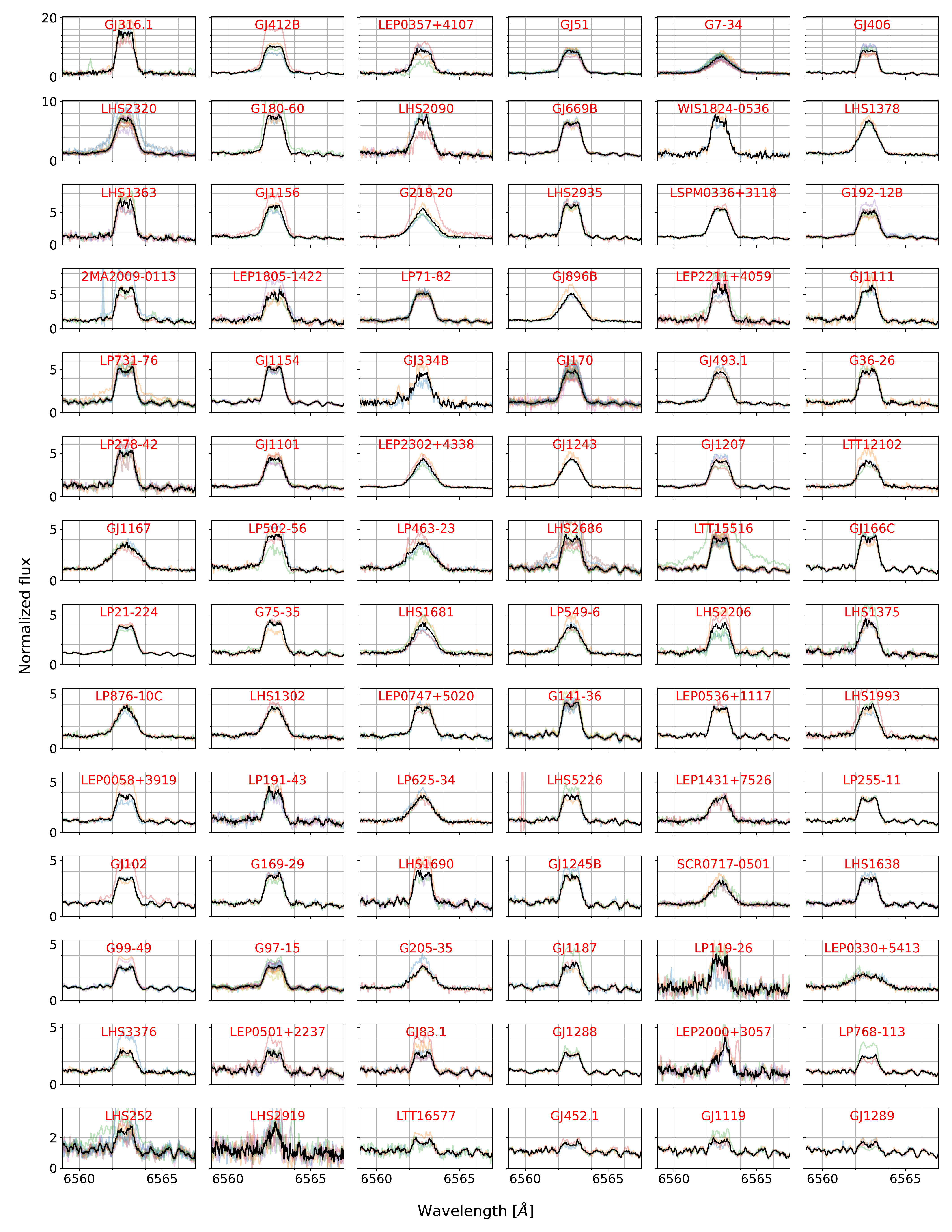}}
    \caption{H\textalpha\ emission for our TRES stars. Individual spectra are shown as colored lines, with the median spectrum in black. Note the y-axis scale varies between rows, with the most active stars shown at the top of the figure.}
    \label{fig:tres}
\end{figure*}

\begin{figure*}[t!]
    \centering
    \makebox[\textwidth][c]{\includegraphics[width=1.\textwidth]{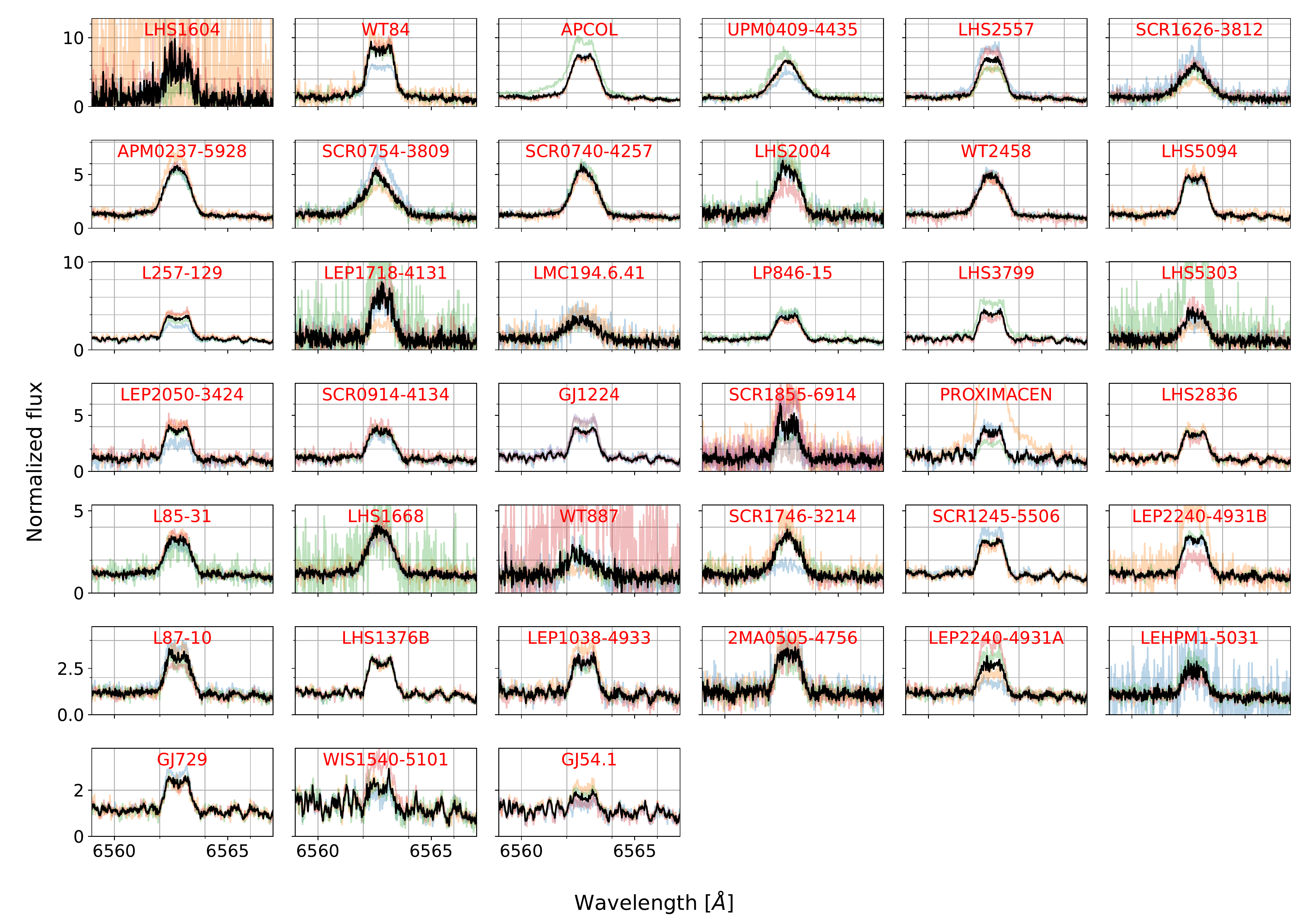}}
    \caption{Same as Figure~\ref{fig:tres}, but for our CHIRON stars.}
    \label{fig:chiron}
\end{figure*}

\section{Data acquisition and reduction}
\label{sec:reduction}
Between 2016 and 2022, we collected multi-epoch, high-resolution observations of each star in the sample using the Tillinghast Reflector Echelle Spectrograph (TRES; $R=44000$) on the \hbox{1.5 m} telescope at the Fred Lawrence Whipple Observatory (FLWO) for sources with $\delta > -15\degree$ and the CTIO HIgh ResolutiON (CHIRON; $R=80000$) spectrograph on the Cerro Tololo Inter-American Observatory (CTIO) \hbox{1.5 m} telescope for sources with $\delta < -15\degree$. The majority of stars were observed at four epochs, which was the objective of the survey. Additional observations were collected for some stars for various reasons, including following up possible RV variability or signatures of binarity.

Our radial-velocity reduction is described in \citet{Pass2023}. In brief, we extract the spectra using the standard TRES \citep{Buchhave2010} and CHIRON \citep{Tokovinin2013} pipelines (which include flat fielding, cosmic-ray rejection, echelle order extraction, and wavelength calibration with ThAr spectra), create a set of templates using coadded observations of stars in the inactive subsample, rotationally broaden these templates, and perform a cross-correlation over wavelength ranges in the regime of 6400--7850\AA. Our pipeline also produces carefully calibrated radial-velocity uncertainties, taking into account the signal-to-noise ratio of the observed spectra, rotational broadening, template mismatch, the long-term stability of the spectrograph, and errors in the barycentric correction. Further details are provided in Section 3.2 of \citet{Pass2023}.

We measure H\textalpha\ equivalent widths following the method defined in \citet{Medina2020}, adopting the convention that a negative value indicates emission. This method uses the wavelength ranges 6554.1--6559.1\AA, 6560.3--6865.3\AA, and 6566.5--6570.5\AA\ for the left continuum, feature, and right continuum, respectively, which were selected to maximize the signal-to-noise in the H\textalpha\ feature while avoiding regions contaminated by telluric lines or molecular bands in M-dwarf spectra. We find that this window size is appropriate for even the most rotationally broadened stars in our sample. Increasing the width of the window from 5\AA\ to 7\AA\ has a negligible impact on our measurements; for LEP 0330+5413, our most rotationally broadened star, this change only results in deviations at the 0.1\AA\ level, much smaller than the variation from spectrum to spectrum.

For each star, we report the median of our equivalent width measurements; while a single measurement may be elevated if we happen to observe the star during a stellar flare, our multi-epoch averages provide a robust estimate of the typical H\textalpha\ activity. In Figure~\ref{fig:tres}, we show a gallery of the H\textalpha\ feature for our TRES targets, with our CHIRON targets in Figure~\ref{fig:chiron}. Note that in the case of a very strong flare, the wings of the H\textalpha\ feature can be enhanced, spilling outside of our measurement window. This effect is not important within the context of this paper, as our median equivalent widths are designed to be uninfluenced by large flares. However, a reader interested in using the epoch H\textalpha\ measurements that we provide should be mindful that a handful of observations with the most extreme emission will be slightly underestimated. This effect is maximized in one observation of LHS 2320, where we report an equivalent width of -18.6\AA. A 7\AA\ window yields -19.6\AA, a deviation of 1\AA.

A summary of our measurements for all 123 stars is given in Table~\ref{tab:summary}, with individual epoch observations presented later in the manuscript (Section~\ref{sec:rv}). To ease readability throughout the text, we have shortened coordinate-based names to their catalog prefix followed by the first four digits of right ascension and declination (e.g., 2MASSJ20091824-0113377 is shortened to \hbox{2MA 2009-0113}); each star's full 2MASS identifier is also given in this table.

\begin{deluxetable}{lccl}[h]
\tabletypesize{\scriptsize}
\tablecaption{Summary of our 123 active stars \label{tab:summary}}
\tablehead{ 
\colhead{Column} & 
\colhead{Format} &  
\colhead{Units} & 
\colhead{Description}} 
\startdata 
1 & A13 & --- & Star name \\
2 & A22 & --- & 2MASS identifier \\
3 & A1 & --- & Instrument (\textbf{T}RES or \textbf{C}HIRON) \\
4 & I2 & --- & Number of spectroscopic observations \\ 
5 & F3.3 & M$_\odot$ & Stellar mass \\
6 & F3.3 & R$_\odot$ & Stellar radius \\
7 & F4.2 & \AA & Median equivalent width of H\textalpha \\
8 & F3.1 & kms$^{-1}$ & Median $v$sin$i$ \\ 
9 & F2.1 & kms$^{-1}$ & $v$sin$i$ standard deviation \\ 
10 & F3.2 & --- & sin$i$ \\ 
11 & F4.3 & kms$^{-1}$ & Median RV uncertainty \\
12 & F4.2 & --- & $\log_{10}(P(\chi^2$)) of a constant model \\ 
13 & F6.3 & days & Rotation period \\
14 & F3.3 & mag & Amplitude of photometric variation \\
15 & I2 & --- & Rotation reference
\enddata
\tablecomments{Full table available in machine-readable form. Detections of rotational broadening less than 3.4 kms$^{-1}$ are consistent with zero given our spectrograph resolution. Rotation references are 1: \citet{Newton2016}; 2: \citet{Newton2018}; 3: \citet{Medina2020}; 4: \citet{Medina2022}; 5: \citet{Pass2022}, 6: \citet{Kiraga2012}; 7: \citet{DiezAlonso2019}, 8: This work, TESS; 9: This work, MEarth; 10: This work, TESS+MEarth. In contrast to some other works, this table reports the peak-to-peak amplitude, not semi-amplitude. For stars for which we measure rotation periods from TESS but the TESS contamination ratio is greater than 1, we do not list an amplitude; while the PDCSAP light curves attempt to correct for contamination from nearby stars \citep{Jenkins2016}, this process is imperfect and we do not consider the amplitudes we measure to be sufficiently reliable when contamination dominates the light curve.}
\end{deluxetable}

\vspace{-0.5cm}
\section{Rotation Periods}
\label{sec:rot}
Activity and rotation are correlated in M dwarfs, such that rapidly rotating stars typically show H\textalpha\ in emission \citep[e.g.,][]{Kiraga2007, Newton2017}. To complement our spectroscopic data, we also report rotation periods for all but eight stars in the sample. We prefer photometric rotation periods based on data from TESS \citep{Ricker2015} and/or the ground-based MEarth array \citep{Nutzman2008, Irwin2015}. Many of these periods were originally published in previous works by the MEarth team \citep{Newton2016, Newton2018, Medina2020, Medina2022, Pass2022}, and we measure 31 new rotation periods using the methods described in those works. In two of these 31 cases, we supplement our analysis with literature rotation periods \citep{Morin2008, Morin2010} to establish which star is the source of which signal, as multiple stars fall within the same TESS pixel; we discuss these systems in Section~\ref{sec:with}. For three stars (GJ 1207, GJ 1224, and GJ 1289), there are insufficient data for a TESS or MEarth detection and we instead report a literature rotation period \citep{Kiraga2012, DiezAlonso2019}.

For the new rotation periods, we note in Table~\ref{tab:summary} whether we measured the period using photometry from TESS, MEarth, or both instruments. We measure rotation periods from MEarth using the method described Section 3 of \citet{Irwin2011} and implemented in the \texttt{sfit} module,\footnote{\href{https://github.com/mdwarfgeek/sfit}{https://github.com/mdwarfgeek/sfit}} which compares a sinusoidal modulation hypothesis to the null hypothesis while accounting for common mode systematics. For TESS, we follow the method described in \citet{Pass2022}, generating Lomb-Scargle periodograms of each source for both the TESS simple aperture photometry (SAP) and pre-search data conditioning SAP (PDCSAP) light curves to ensure we are not misidentifying systematics as rotational modulation, nor is rotational modulation being mistakenly removed as systematics. After visually confirming the existence and period of rotational modulation, we measure its amplitude using a modified version of \texttt{sfit}.

\subsection{Stars without rotation periods}
In this section, we discuss the eight stars without measured rotation periods (either from TESS/MEarth or from another literature source). Note that there are multiple reasons why we might fail to obtain a rotation period, including an absence of data, poor data due to contamination by a nearby bright star, the amplitude of the modulation being too small to detect, or the period being too long to detect given the observation baseline. The latter effect is particularly important for TESS, as periods longer than 10 days are difficult to measure with TESS due to the 27-day duration of each sector.
\label{sec:without}

\textbf{GJ 166 C} has been observed in one TESS sector at 2-minute cadence, from which we cannot measure a rotation period. The A component of this system is a K-dwarf at 78" separation, which contaminates the TESS aperture. We are also unable to measure a period from MEarth data. Our measurement of $v$sin$i$, 1.7 kms$^{-1}$, is consistent with no broadening at our spectrograph resolution, but the star has a large H\textalpha\ equivalent width of -4.4\AA. This activity may be the result of the recent evolution of the system's B component into a white dwarf (see discussion in \citealt{Fuhrmann2014} and \citealt{Pass2022}).

\textbf{LP 119-26} has been observed in two TESS sectors at 2-minute cadence. The light curve possibly exhibits a weak 5.8-day periodicity, although we are not confident enough in the reality of this signal to claim it as a detection. \citet{Newton2016} identified an uncertain (`U-grade') rotation period of 6.214 days, promisingly consistent with this candidate period, although additional MEarth data observed since the publication of that work do not resolve this uncertainty. This star is one of the faintest in our sample. It is not rotationally broadened at the resolution of the spectrograph (which is consistent with a 6-day period) and is moderately active, with a median H\textalpha\ equivalent width of -2.7\AA. While \citet{Gagne2018a} identified this star as a candidate member of the 200Myr-old Carina-Near moving group, their assessment was made in the absence of a radial velocity measurement. With our measured radial velocity, the BANYAN $\Sigma$ tool \citep{Gagne2018} yields a negligible chance that this star is a member of Carina-Near.

\textbf{LP 731-76} has been observed in two TESS sectors at 2-minute cadence, from which we cannot measure a rotation period. The TESS light curve is contaminated by a nearby, unassociated K dwarf, with a contamination ratio of 4.9 according to the TESS Input Catalog (TIC; \citealt{Stassun2019}). This ratio indicates that the light curve contains nearly five times as much flux from contaminating stars as it does flux from LP 731-76. There are also insufficient data for a MEarth period determination. Our $v$sin$i$ measurement of 2.5kms$^{-1}$ is consistent with no broadening given the resolution of the spectrograph, but the star is highly active, with a median H\textalpha\ equivalent width of -5.5\AA.

\textbf{GJ 412 B} has been observed in one TESS sector, but not at 2-minute cadence. We are unable to determine a rotation period from the TESS full-frame image or the MEarth data. The TESS data are contaminated by the much brighter early M companion at 32" separation, GJ 412 A. With our measured $v$sin$i$ of 4.5 kms$^{-1}$, GJ 412 B should have a short rotation period of less than 1.8 days. This star is highly active, with a median H\textalpha\ equivalent width of -13.5\AA.

\textbf{LHS 2919} has been observed in one TESS sector at 2-minute cadence, from which we cannot measure a rotation period. We also cannot identify a rotation period in the extant MEarth data. There are large gaps in the TESS light curve due to scattered light; the incompleteness of the light curve may contribute to our null detection. This star is also one of the faintest in our sample. We measure a $v$sin$i$ of 3.9 kms$^{-1}$, suggesting a rotation period of 2 days or less. The median H\textalpha\ equivalent width is -1.2\AA, near the limit of what we consider to be active.

\begin{figure}[t]
    \centering
    \includegraphics[width=\columnwidth]{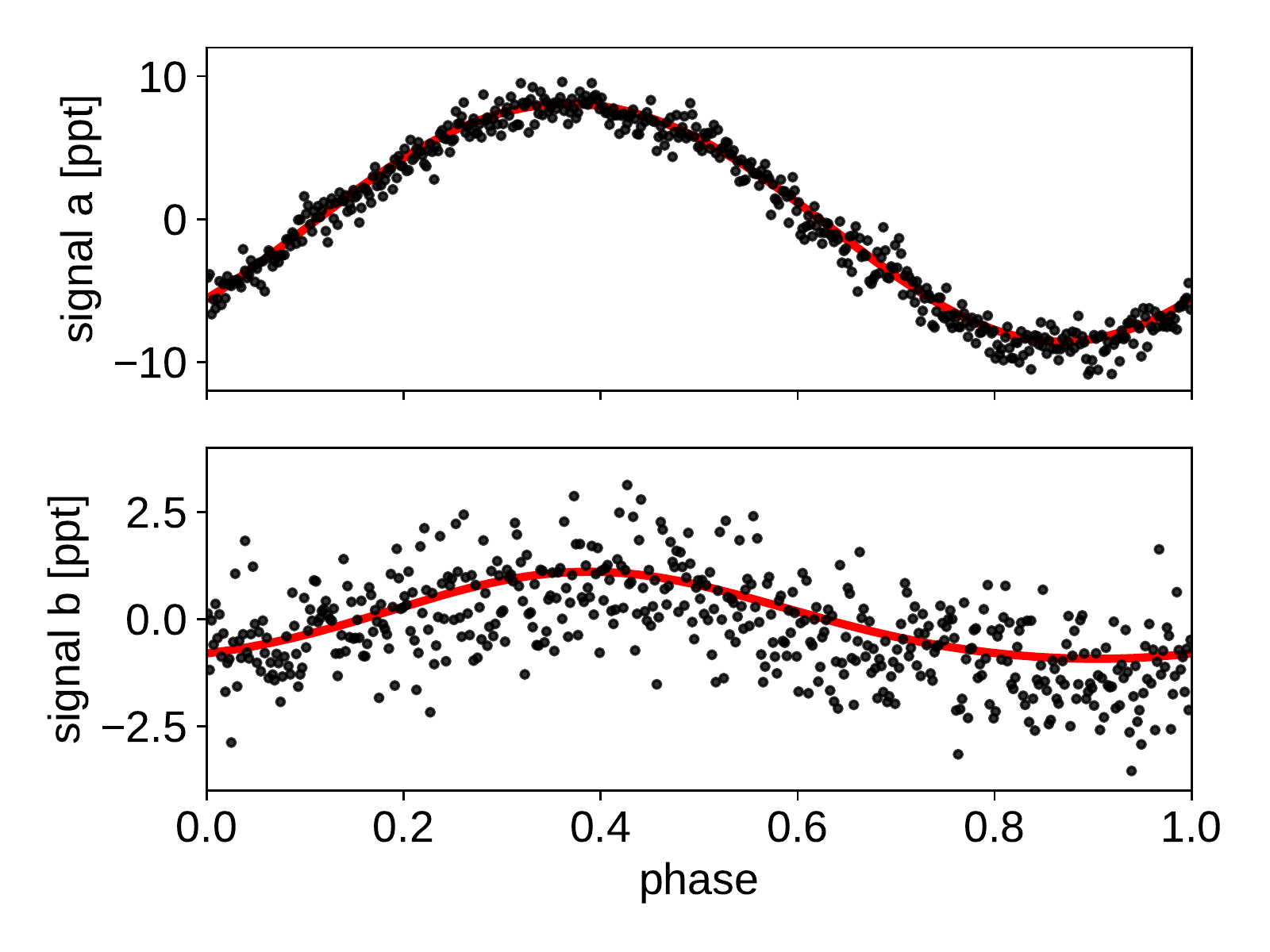}
    \caption{The phased sector 1 PDCSAP light curve for TIC 161174284, containing LEP 2240-4931 AB. The top panel shows the 1.002-day signal (with the 0.598-day model removed) while the bottom panel shows the smaller 0.598-day signal (with the 1.002-day model removed). The points show 500 bins evenly spaced in phase. The best-fitting spot model is in red, which includes sinusoids for the fundamental mode and the first harmonic.}
    \label{fig:LEP2240}
\end{figure}

\begin{figure}[t]
    \centering
    \hspace{-1cm}
    \includegraphics[width=1.1\columnwidth]{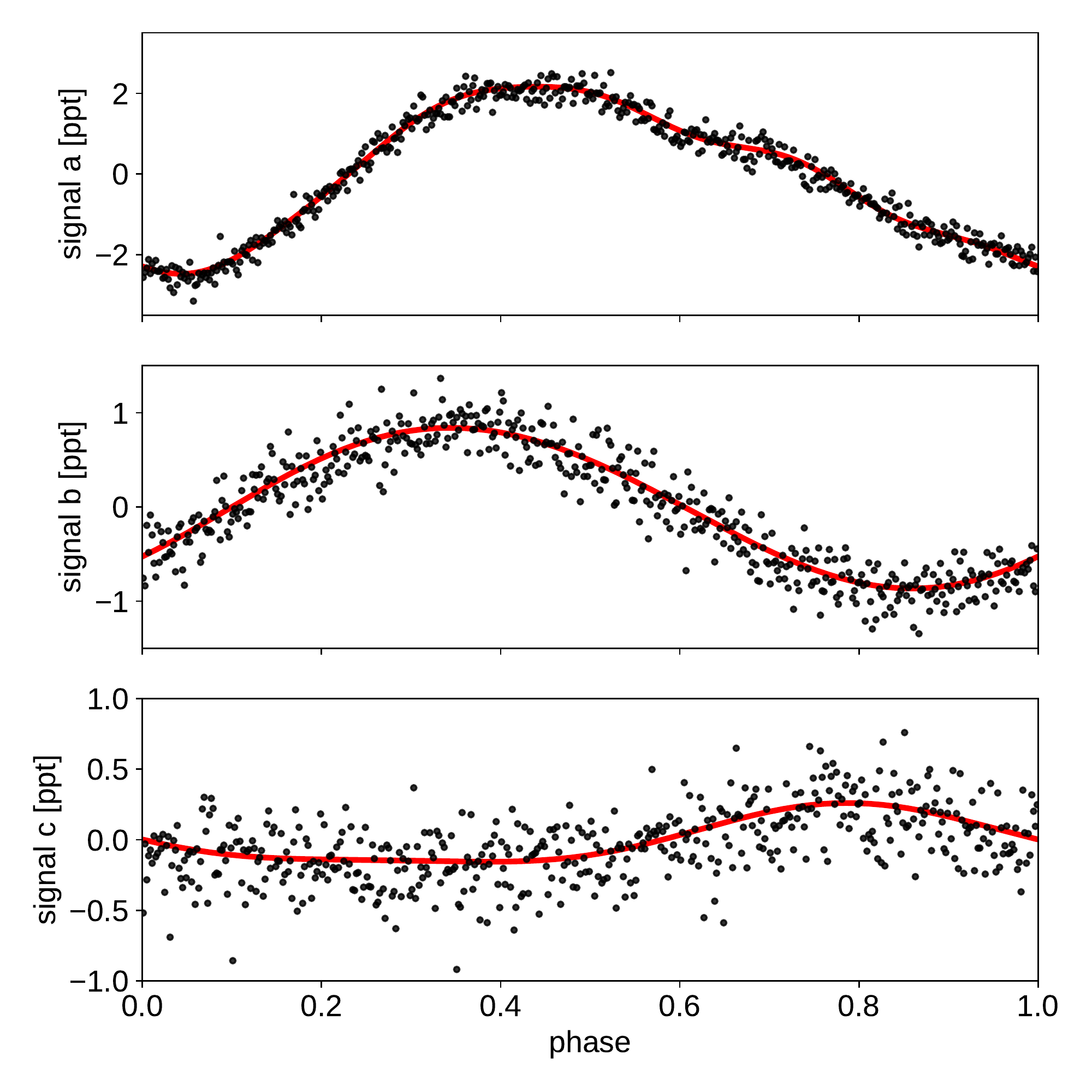}
    \caption{The phased sector 35-37 PDCSAP light curve for TIC 342888849, containing LHS 2004 and LHS 2005 AB. This target was also observed in sectors 8-10, exhibiting similar signals, but the spot pattern evolves sufficiently over two years that we omit the earlier sectors from the plot for clarity. The top panel shows the 1.174-day signal, the middle shows the 0.139-day signal, and the bottom shows the 0.210-day signal. In each case, the models of the other two signals have been removed. The points show 500 bins evenly spaced in phase. The best-fitting spot model is in red. To sufficiently model the detail in signal a, we include sinusoids for the fundamental mode and four harmonics. We only require the fundamental mode and the first harmonic for b and c.}
    \label{fig:LHS2004}
\end{figure}

\textbf{LEP 1805-1422} has not been observed with TESS, and is not scheduled to be observed with TESS through sector 69. We do not detect a rotation period in the observations available from MEarth. With a measured $v$sin$i$ of 5.8 kms$^{-1}$, this star should have a short rotation period of less than 1.5 days. This star is highly active, with a median H\textalpha\ equivalent width of -5.9\AA.

\textbf{LEP 2240-4931 A \& B} fall within the same TESS pixel. The TESS light curve shows a strong 1.002-day period, with a small 0.598-day residual (Figure~\ref{fig:LEP2240}). It is unclear which component is responsible for which signal. MEarth is able to resolve the two components, but we do not identify rotation periods in the MEarth data for either star. While the 1.002-day signal is strong in TESS, such a period would be difficult to extract from ground-based light curves due its proximity to the 1-day alias. The 0.598-day signal would also be difficult to observe with MEarth due to its small amplitude. The stars have similar rotational broadening (median $v$sin$i$ of 7.5 kms$^{-1}$ and 8.3 kms$^{-1}$) and activity levels (median H\textalpha\ equivalent widths of -2.1\AA\ and -2.9\AA).

\textbf{Future prospects:} For the six outstanding stars with declinations above -15\degree, a campaign is underway to obtain rotation periods using the Tierras Observatory \citep{GarciaMejia2020}, an ultra-precise photometer optimized for the study of M dwarfs.

\needspace{6em}
\subsection{Stars with rotation periods}
\label{sec:with}
There are also a few stars with measured rotation periods that merit discussion.

\textbf{LP 768-113} and \textbf{LEP 0058+3919} have rotation periods of 5.07 days and 0.457 days as published in \hbox{\citet{Medina2020}} and \citet{Medina2022}, respectively. We present revised rotation period estimates for these stars. For LEP 0058+3919, our analysis indicates that the 0.457-day signal is a harmonic of the true rotation period at 0.914 days. For LP 768-113, we measure a somewhat longer rotation period of 7.7 days using two sectors of 2-minute cadence TESS data. Such a long rotation period is challenging to measure with TESS, as the TESS PDCSAP light curves can remove real astrophysical signals at these periods, alongside the detector systematics. Notably, \citet{Medina2020} did not measure any TESS rotation periods longer than 6.0 days. Here we use the SAP light curves and remove the systematics ourselves, as described in \citet{Pass2022}, allowing us to mitigate this issue. We also detect a period of 7.853 days from the MEarth data, supporting our conclusions.

\textbf{LHS 2004} is a member of a triple system, separated by 8" from the close binary LHS 2005 AB. All three stars therefore fall into the same TESS pixel. We observe periodogram peaks consistent with three rotation periods in TESS, alongside their harmonics: 1.174 days, 0.139 days, and 0.210 days, in descending order of strength (Figure~\ref{fig:LHS2004}). We observe both the 1.174-day and 0.139-day signals in the MEarth data of LHS 2005 AB, suggesting that the 0.210-day period belongs to LHS 2004, which we adopt here; however, the MEarth data of LHS 2004 are insufficient for us to measure this (or another) rotation period from them directly.

\textbf{GJ 1245 B} is a member of a triple system, separated by 6" from the close binary GJ 1245 AC. All three stars therefore fall into the same 21" square TESS pixel. We measure 0.709-day and 0.263-day signals in the blended TESS light curve, as well as harmonics of these signals. There are not sufficient MEarth data to determine which period corresponds to which star. However, \citet{Hartman2011} report a rotation period of 0.26 days for either A or C based on photometric monitoring by HATNet, and \citet{Morin2010} report a rotation period of 0.710 days for B based on a Zeeman-Doppler imaging analysis of a time series of circularly polarized spectra. The TESS data therefore support a 0.709-day rotation period for GJ 1245 B.

\textbf{GJ 896 B} is a part of a binary, separated by 5" from the earlier M dwarf GJ 896 A. The blended TESS light curve shows a strong 1.067-day periodicity as well as a small 0.404-day residual. Based on Zeeman-Doppler imaging analysis of a time series of circularly polarized spectra, \citet{Morin2008} report a rotation period of 1.061 days for A and 0.404 days for B. We therefore associate the 0.404 TESS signal with GJ 896 B.

\subsection{Single stars with multiple rotation periods}
For a handful of presumed-single stars, we detect a second candidate rotation period in the TESS residuals. We note those detections here. Such signals could be due to an unresolved companion, or, as the TESS pixels are large, could be a contaminating signal from a background star. Note that the median TESS contamination ratio for stars in our sample is 0.03; all of the stars with a second candidate signal have a substantially larger contamination ratio than this median, supporting the hypothesis that a background star is responsible. Moreover, one of us (J.\ Winters) has gathered unpublished speckle observations that rule out a companion for the first four of these targets, and the POKEMON speckle survey \citep{Clark2022} did not find any companions for the fifth star (C.\ Clark, private communication).  We also do not detect statistically significant RV variations for any of these stars that would indicate the presence of a close companion, nor do any of these stars have a high Gaia Renormalised Unit Weight Error (RUWE) that would suggest an astrometric perturber.

For \textbf{WIS 1540-5101} we measure a weak 0.165-day signature in TESS; the rotation period of this star was measured to be 93.702 days using MEarth in \citet{Newton2018}. The TIC lists the contamination ratio as 0.49, meaning roughly a third of the light in the aperture is from contaminating sources.

For \textbf{L 257-129}, we measure an 11.891-day rotation period with MEarth but note a strong 0.253-day residual in TESS. This star has a TESS contamination ratio of 0.67, meaning about 40\% of the light in the aperture is from contaminating sources. The 0.253-day signal is also observable in the Quick Look Pipeline (QLP; \citealt{Huang2020}) light curves of other nearby stars, including TIC 256911699 and TIC 256911665.

For \textbf{LEP 1718-4131}, we measure a 1.516-day period in TESS and see peaks consistent with this period in MEarth, but also observe a 0.625-day residual in TESS, along with harmonics of this period. We do not see this second signal in MEarth. This star has a TESS contamination ratio of 1.07, meaning the contaminants contribute more light than does LEP 1718-4131.

\citet{Medina2020} report a 0.70-day period for \textbf{SCR 1245-5506}, which we recover in TESS and MEarth, although we also note a TESS residual at 0.311 days. The TESS contamination ratio is 0.14 and we see the 0.311-day signal in the QLP light curve of nearby stars such as TIC 419692043.

For \textbf{LTT 12102}, \citet{Newton2016} report a rotation period of 0.576 days from MEarth, which we recover with TESS, although we also note a 0.807-day residual in the TESS periodogram. The TESS contamination ratio is 0.23 and we also see the signal in nearby stars such as TIC 271204415.

\subsection{Discussion of rotation periods}
Of the 123 stars in our volume-complete sample, 86 have rotation periods shorter than 2 days, 22 have rotation periods between 2 and 20 days, 9 have rotation periods longer than 20 days, and 6 have undetermined rotation periods. Considering binomial uncertainties for the 117 stars with measured rotation periods, our results indicate that 74$\pm$4\% of active, low-mass M dwarfs rotate with periods shorter than 2 days, 19$\pm$4\% with periods of 2--20 days, and 8$\pm$2\% with periods longer than 20 days. Depending on the nature of the 6 unclassified stars, the central value of these estimates could vary from 70--75\%, 18--23\%, and 7--12\%, respectively. Based on our $v$sin$i$ measurements, we argue in Section~\ref{sec:without} that at least three of the stars with undetermined rotation periods rotate with periods shorter than 2 days, tightening these ranges to 72--75\%, 18--20\%, and 7--10\%, respectively, or 72--74\%, 19--20\%, and 7--9\% if we also adopt the 6-day candidate period for LP 119-26. Binomial uncertainty therefore dominates the error budget in all three bins, with the uncertainty in the longest-period bin rounding up to 3\% and the others remaining unchanged.

\begin{figure}[t]
    \centering
    \makebox[\columnwidth][c]{\includegraphics[width=1.05\columnwidth]{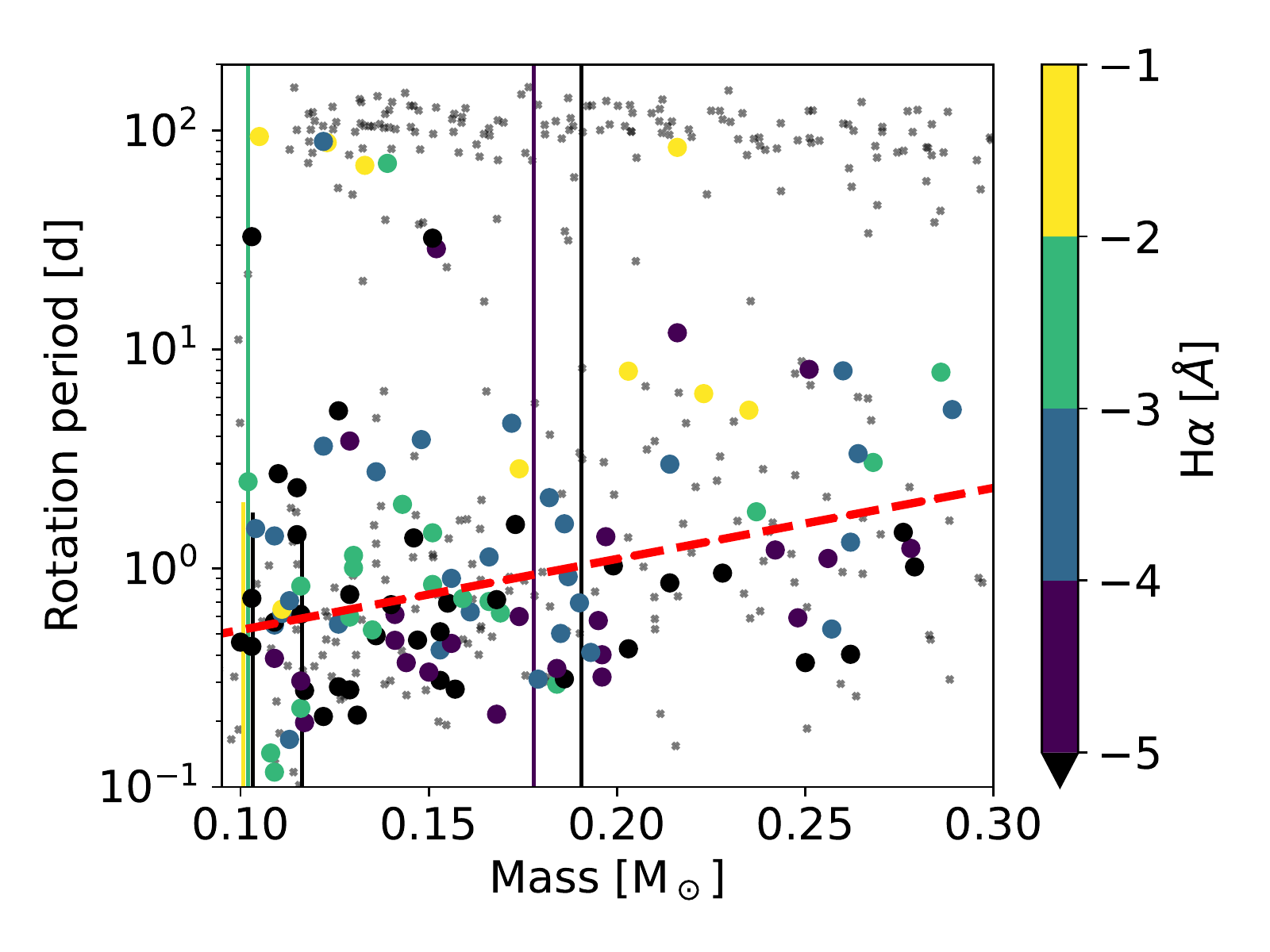}}
    \caption{The mass--rotation diagram for our active sample. The six stars without measured rotation periods are noted by vertical lines (we are able to include LEP 2240-4931 AB in the scatter plot despite not knowing which component is responsible for which period, as the pair is an equal-mass binary). For three of the six, we place an upper limit on the rotation period using our $v$sin$i$ measurement. The red dashed line represents a linear regression over all stars with periods less than 20 days, suggesting a weak positive correlation between the rotation periods of rapidly rotating M dwarfs and mass. This line is defined by the equation $\log_{10}(P_{\rm rot})=-0.61 + 3.2M_*$, with $P_{\rm rot}$ in days and $M_*$ in solar masses. For comparison, the sample from \citet{Newton2016, Newton2018} is shown as smaller points in the background; that sample includes both active and inactive stars.}
    \label{fig:mprot}
\end{figure}

Figure~\ref{fig:mprot} presents these rotation periods as a function of stellar mass. \citet{Newton2017} showed that M dwarfs exhibit bimodal rotation periods: active stars appear as a rapidly rotating population while inactive stars form a slowly rotating sequence. As our sample is selected based on its H\textalpha\ activity, the majority of our stars are in the rapidly rotating mode, which we find exhibits a mass dependence. The mass dependence for rapid rotators is opposite in sign to the trend found for slow rotators in \citet{Newton2017}. That is to say, lower-mass M dwarfs tend to rotate faster than more massive M dwarfs when in the rapidly rotating mode, but more slowly than their massive counterparts once they have spun down. Such a mass dependence for rapidly rotating M dwarfs has been previously observed in young clusters, with \citet{Somers2017} arguing that the trend is an artifact of physical processes imprinted during the pre-main-sequence phase.

A small number of active stars have spun down to long rotation periods, appearing in a similar portion of the mass--rotation diagram as the inactive, slowly rotating sequence, as can be seen when comparing with the \citet{Newton2016, Newton2018} sample in Figure~\ref{fig:mprot}. However, these spun-down active stars all fall on the shorter-rotation side of this group of slowly rotating points, perhaps suggesting these stars are newly spun down. This group includes Proxima Centauri, our nearest neighbor, which has a rotation period of 89 days and an H\textalpha\ equivalent width of -3.3\AA.

\begin{figure}[t]
    \centering
    \makebox[\columnwidth][c]{\includegraphics[width=\columnwidth]{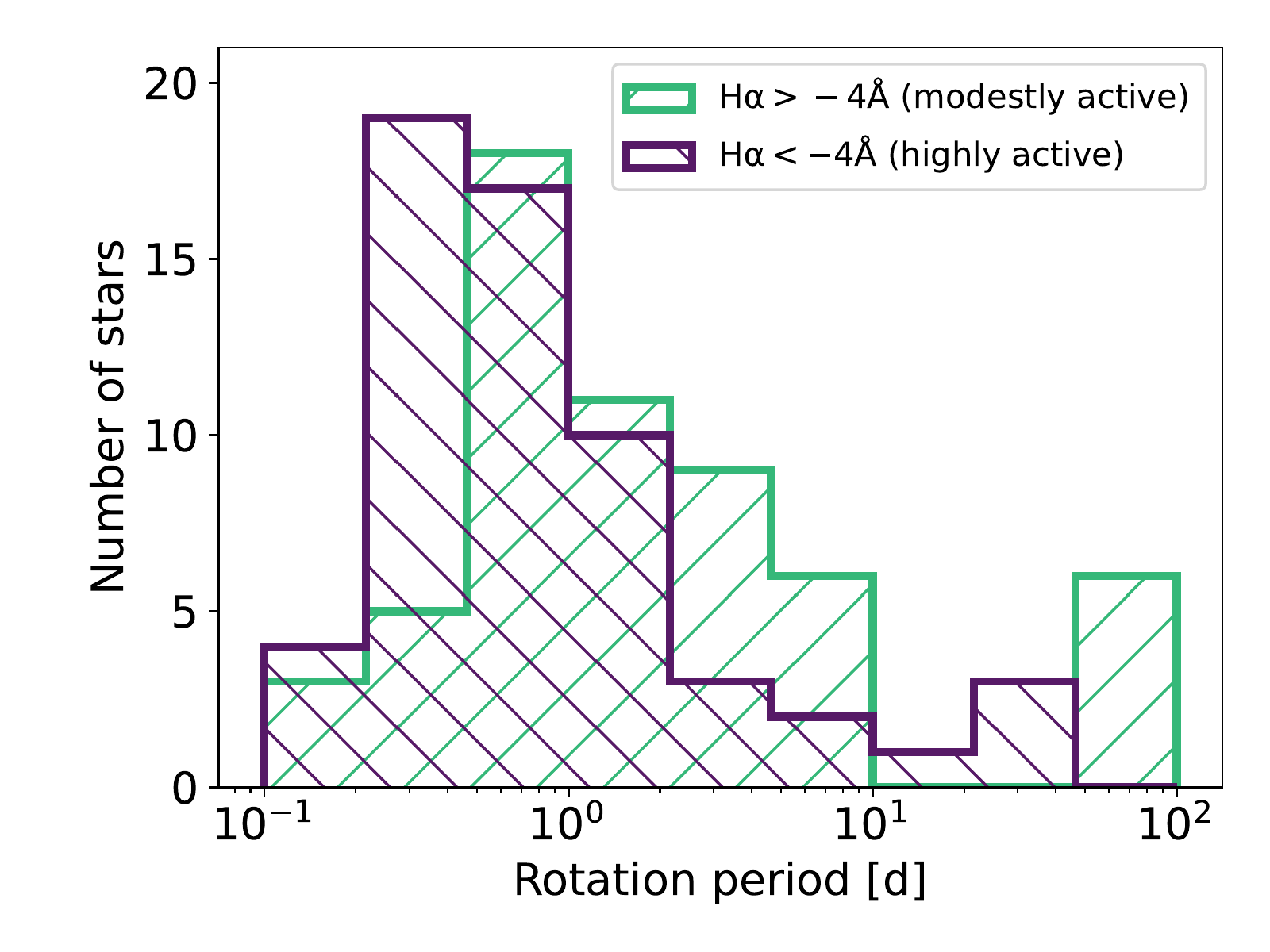}}
    \caption{A simplified version of Figure~\ref{fig:mprot}. Here we split our sample into two equally sized groups: stars that are highly active (\hbox{H\textalpha\ $< -4$\AA,} shown in purple) and those that are more modestly active (\hbox{H\textalpha\ $> -4$\AA,} shown in green). Within the rapidly rotating mode, the highly active stars tend to have shorter rotation periods and the modestly active stars tend to have longer rotation periods.}
    \label{fig:hist}
\end{figure}

There is not a clear trend between H\textalpha\ equivalent width and rotation period within the active population; this phenomenon defines the ``saturated regime" reported in previous works \citep[e.g.,][]{Newton2017}. However, there is still some correlation. In Figure~\ref{fig:hist}, we show that very rapidly rotating M dwarfs ($P_{\rm rot} < 0.5$ days) tend to have very high levels of H\textalpha\ emission (equivalent widths beyond \hbox{-4\AA}) and the more slowly rotating stars in the rapidly rotating mode ($2 < P_{\rm rot} < 10$ days) tend to have more modest H\textalpha\ emission (shallower than \hbox{-4\AA}). We use -4\AA\ as the division between modestly active and highly active because it divides the sample into two equally sized halves. A two-sample Kolmogorov-Smirnov (KS) test indicates that the difference between the distributions is statistically significant, with a p-value of 0.0035. As we discuss in \citet{Pass2022}, the older stars in the rapidly rotating reservoir tend to have longer ($2 < P_{\rm rot} < 10$-day) rotation periods. The offset in Figure~\ref{fig:hist} may therefore be the result of H\textalpha\ emission tempering with age, perhaps reflecting a changing magnetic field complexity \citep[e.g.,][]{Garraffo2018}. Galactic kinematics also suggests that the two groups have different ages: using the velocity dispersion method described in \citet{Medina2022} and based on \citet{Lu2021}, we estimate a characteristic age for the highly active population that is nearly a gigayear younger than the modestly active population (with an age of around 2 Gyr for the highly active population and 3 Gyr for the modestly active population). This technique relates the age of a stellar sample to its velocity dispersion in the direction of the Galactic north pole. Restricting the sample to only stars with rotation periods shorter than 20 days does not have a significant impact on these results. 

\begin{figure}[t]
    \centering
    \makebox[\columnwidth][c]{\includegraphics[width=\columnwidth]{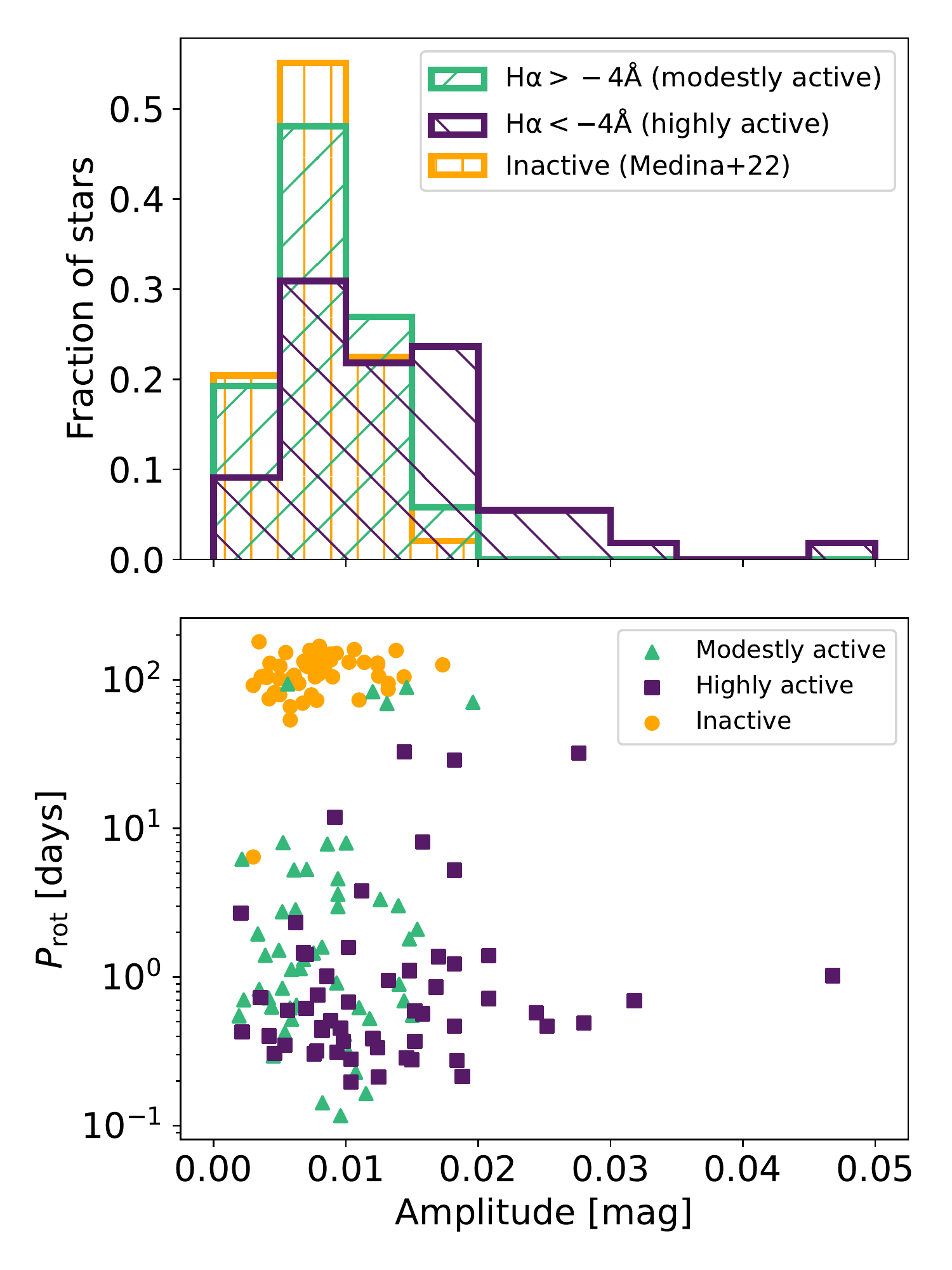}}
    \caption{In the upper panel, we show a histogram of the photometric peak-to-peak amplitudes of stars with $P_{\rm rot}$ measured from TESS or MEarth, separated into the same two H\textalpha\ bins as in Figure~\ref{fig:hist}. In orange, we show inactive \hbox{(H\textalpha\ $> -1$\AA)} stars with $P_{\rm rot}$ from TESS or MEarth, using amplitudes tabulated in \citet{Medina2022}. Like the active stars, these inactive stars are part of the 15pc sample of 0.1--0.3M$_\odot$ M dwarfs and have been vetted for close binary companions. The lower panel shows these same data in the amplitude--rotation-period plane. Highly active stars tend to have larger amplitudes than modestly active or inactive stars.}
    \label{fig:amp}
\end{figure}

Stars in the highly active population also tend to exhibit larger photometric amplitudes than stars in the modestly active population, as shown in Figure~\ref{fig:amp}. A KS test yields a p-value of 0.005; i.e., there is only a 0.5\% chance that the amplitudes of the very active sample and the modestly active sample are drawn from the same distribution. The difference between these distributions lies in the large-amplitude tail: stars with photometric amplitudes larger than 0.015 mag are much more likely to be highly active. On the other hand, the distribution of amplitudes for the modestly active stars is not statistically distinct from the distribution for inactive stars: when comparing with the inactive sample from \citet{Medina2022}, we find a p-value of 0.24. The existence of a correlation between H\textalpha\ activity and photometric amplitude is in contrast to \citet{Newton2016} and \citet{Medina2022}, who found no statistically significant correlation between the rotation periods of low-mass M dwarfs and their observed photometric amplitudes. However, neither work specifically considered whether there was a difference between the highly and modestly H\textalpha-active populations, as we have here.

Another hint of a correlation is that the stars from our active sample with the least H\textalpha\ emission (-1 to -2\AA, shown in yellow in Figure~\ref{fig:mprot}) have longer rotation periods on average. That said, some of the stars with rotation periods longer than 20 days have very high levels of H\textalpha\ emission; in particular, there are three stars with rotation periods of 20--40 days and masses less than 0.2M$_\odot$, each exhibiting H\textalpha\ emission in excess of -4\AA. \hbox{GJ 316.1} is one of these stars and has the most H\textalpha\ emission of all stars in the survey, with an equivalent width of \hbox{-17.2\AA.} While we only have three stars in this region of parameter space in our sample, stars in a similar location on the mass--rotation diagram from the \citet{Newton2016, Newton2018} sample show similar properties. That sample contains eight other stars that fall within the 20 $<$ $P_{\rm rot}$ $<$ 40 days, $M_* < 0.2$M$_\odot$ bin, with masses revised using Gaia parallaxes and the \citet{Benedict2016} $K$-band mass--luminosity relation. Five of those stars had their H\textalpha\ emission measured in \citet{Newton2017}, with LHS 2243 and LP 373-35 showing some of the highest H\textalpha\ activity levels in that survey, with equivalent widths of -26.8\AA\ and -17.6\AA. LP 524-48 and LP 604-16 also show very high H\textalpha\ emission levels of -6.6\AA\ and -9.1\AA. Only \hbox{G 141-53} exhibits modest H\textalpha\ emission, with an equivalent width of -2.5\AA. As these stars fall within the gap between the rapidly rotating and slowly rotating modes, we suggest that these stars are currently experiencing rapid spindown, and their high H\textalpha\ emission levels may result from the spindown processes responsible for their rapid loss of angular momentum. Two of the three gap stars from our survey were studied in \citet{Medina2022}, who found that these stars had rates of flaring that were higher than the average star in the saturated regime; these rates were comparable to the M dwarfs in young moving groups studied in that work. \citet{Mondrik2019} previously posited that M dwarfs with intermediate rotation periods may have enhanced flare rates due to changing magnetic field geometries. An alternate hypothesis is that these gap stars are among the group we identified in \citet{Pass2022} that spin down to the slowly rotating sequence at younger ages than the average M dwarf. However, this would not explain why there are no modest-activity M dwarfs in the gap, unless stars that spin down young traverse the gap more slowly than stars that spin down at older ages.

\section{Inclinations}
\label{sec:i}
With a photometric rotation period and a spectroscopic $v$sin$i$, we are able to constrain a star's inclination. This requires the equation of circular motion, $v = 2\pi R_* / P_{\rm rot}$, and neglects differential rotation; however, differential rotation is expected to be negligible for rapidly rotating, low-mass M dwarfs (see Section 7.1 of \citealt{Kesseli2018} and references therein). We estimate $R_*$ for our stars using the mass--radius relation derived using interferometric radii in \citet[][their equation 10]{Boyajian2012}, with stellar masses estimated from the \citet{Benedict2016} $K$-band relation and tabulated in \citet{Winters2021}.

To use this method of estimating $R_*$, we must trust that our absolute $K$-band magnitudes are accurate. We note firstly that all of our stars have precise Gaia parallaxes; therefore, the conversion between apparent and absolute magnitude does not introduce notable uncertainty. Secondly, we have neglected close binary stars from our sample, with our multi-epoch, high-resolution spectroscopic observations allowing us to detect previously unknown unresolved binaries \citep[][and further discoveries in prep]{Winters2018, Winters2020}. We therefore expect the 2MASS $K$-band magnitudes of our stars to be generally free from contamination. We must also trust that the \citet{Boyajian2012} relation is accurate for our stars. While past works have posited a link between rapid rotation and radius inflation \citep[e.g.,][]{Kraus2011}, \citet{Kesseli2018} established that rapid rotation does not inflate the radii of fully convective M dwarfs, and specifically, they verified that the \citealt{Benedict2016}+\citealt{Boyajian2012} method employed here results in radii for rapidly rotating, magnetically active, fully convective M dwarfs that are accurate to within 5\% errors. We therefore assert that this method will produce reasonable estimates of $R_*$ for the purposes of our inclination analysis.

\begin{figure}[t]
    \centering
    \includegraphics[width=\columnwidth]{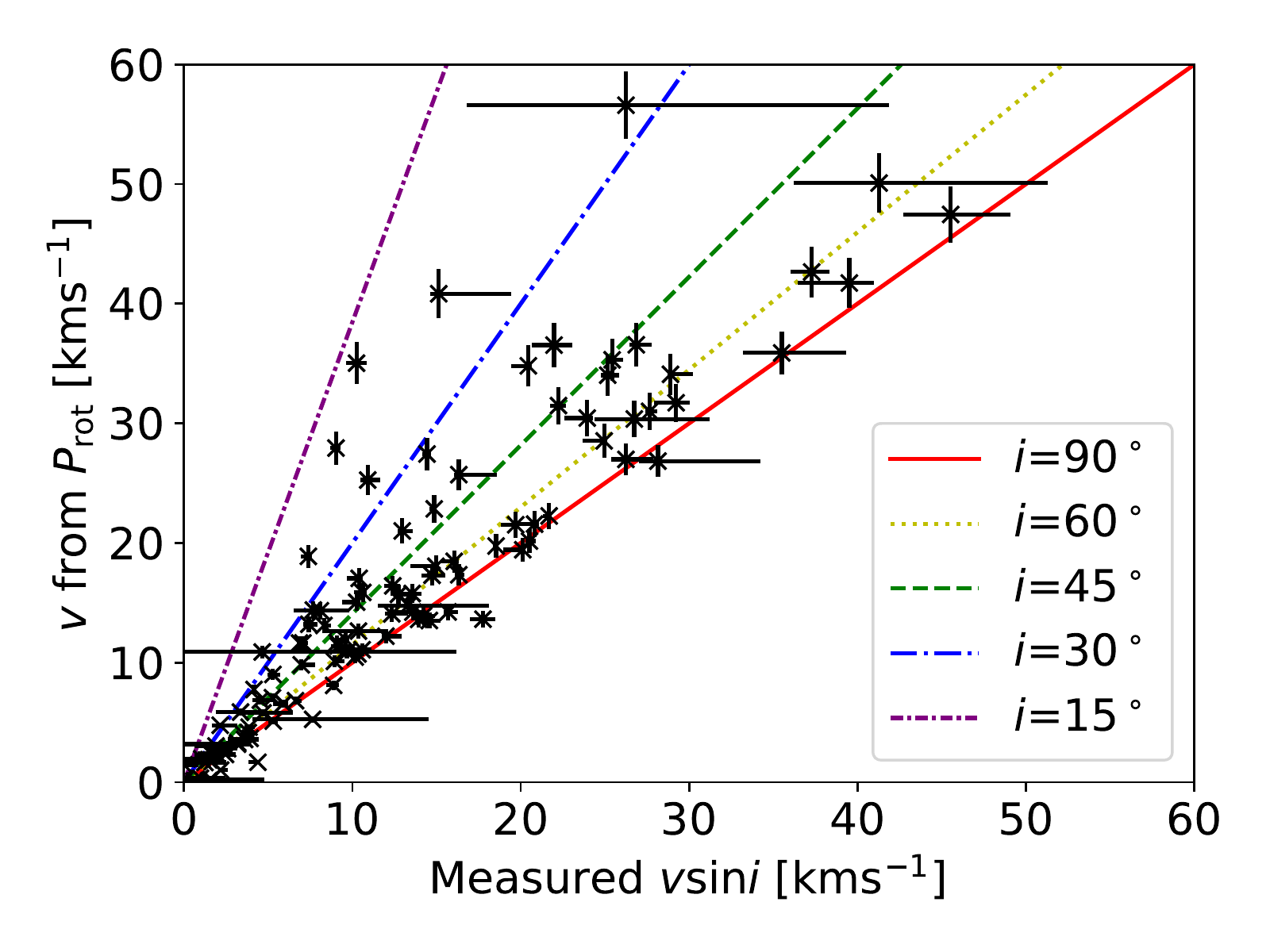}
    \caption{The spectroscopic $v$sin$i$ and photometric $v$ for our 115 stars with measured rotation periods. Detections of rotational broadening less than 3.4 kms$^{-1}$ are consistent with zero given our spectrograph resolution. Colored lines indicate constant inclination. The y-axis error bars illustrate 5\% uncertainties on the radii and the x-axis error bars show the range of $v$sin$i$ measurements observed for each star. No stars should fall below the red \hbox{$i=90$\degree} line, although we observe a small number of interlopers in this region. We discuss possible reasons for these outliers in the text.}
    \label{fig:inclinations}
\end{figure}

\begin{figure}[t]
    \centering
    \includegraphics[width=\columnwidth]{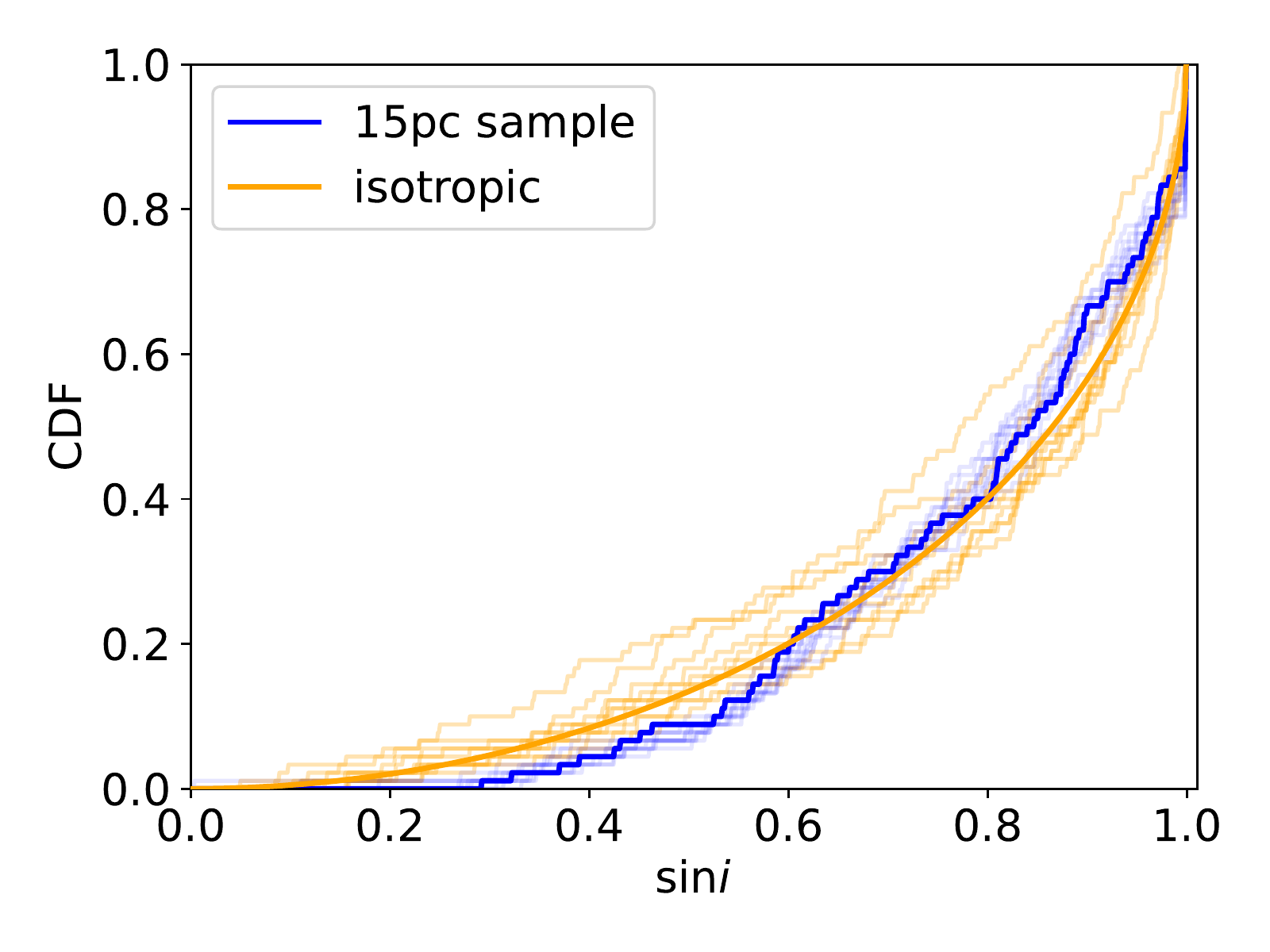}
    \caption{In solid blue, we show the cumulative distribution of inclinations for the 90 stars with $v > 3.4$ kms$^{-1}$. In transparent blue, we show ten random draws consistent with the uncertainties in our observed parameters; we assume 5\% uncertainties on the radii and a uniform range of $v$sin$i$ between the minimum value and the maximum value we observe for each star. The solid orange line shows an isotropic distribution of spin axes. In transparent orange, we show ten random draws from the isotropic distribution given our sample size.}
    \label{fig:iso}
\end{figure}

In Figure~\ref{fig:inclinations}, we compare our spectroscopic $v$sin$i$ to the velocity $v$ estimated from $P_{\rm rot}$. None of our stars with measured rotation periods suggest inclinations more pole-on than 15\degree. Assuming an isotropic distribution of spin axes, we would expect $1 - \cos(15$\degree$)=3.4\%$ of our stars to have inclinations below 15\degree, or roughly 3 out of the 90 stars with $v > 3.4$ kms$^{-1}$ (for stars with $v < 3.4$ kms$^{-1}$, we expect to observe a $v$sin$i$ consistent with zero regardless of the value of $i$, as we are limited by our spectrograph resolution). Moreover, there is a 96\% chance we would have detected at least one star with an inclination below 15\degree. This may indicate that the pole-on stars are in the small group without measured photometric rotation periods. This explanation is sensible: a pole-on orientation means that the face of the star oriented towards us does not change greatly as a function of phase, resulting in a decreased amplitude of photometric variability. Exempting our lack of pole-on stars, the distribution of inclinations is well described by the isotropic model (Figure~\ref{fig:iso}). Note that there are two stars with $v > 3.4$ kms$^{-1}$ but $v$sin$i < 3.4$ kms$^{-1}$: WIS 1824-0536, with $v=5.9$ kms$^{-1}$ and GJ 83.1, with $v=4.8$ kms$^{-1}$. It cannot be ruled out that one or both of these stars is pole-on, as we can only establish that $i$ must be less than 35\degree\ and 45\degree, respectively. However, we have included both these stars in Figure~\ref{fig:iso} using their nominally measured $v$sin$i$, which in both cases corresponds to sin$i$ less than 0.6. Therefore, the underdensity observed in the cumulative distribution function (CDF) for sin$i < 0.6$ persists even if the $v$sin$i$ of these two stars are adjusted to lower values. That is to say, our lack of pole-on stars is not simply an artifact of our spectrograph resolution.

Thirteen of the 90 stars have sin$i > 1$, which is unphysical; for these stars, we have capped sin$i$ at 1. In most cases, these values do not exceed 1.1 and are likely the result of modest uncertainties in our measurements of $v$sin$i$, $P_{\rm rot}$, and $R_*$. However, two are more significant outliers: LEP 1718-4131, with sin$i=1.46$, and \hbox{GJ 334 B}, with sin$i=1.31$. For LEP 1718-4131, we suspect the issue is a large uncertainty in our measurement of $v$sin$i$. As this star is faint, we attain a low cross-correlation coefficient of roughly $h=0.3$ in each of our four CHIRON spectroscopic observations, and our estimate of $v$sin$i$ varies greatly between spectra: 4.1, 11.0, 4.2, and 14.5 kms$^{-1}$, resulting in the median value of 7.6 kms$^{-1}$ that we have adopted to calculate sin$i$. If we instead use 4.1 kms$^{-1}$, we find sin$i=0.78$, a physically reasonable result. Attempting to jointly fit the four spectra with the same value of $v$sin$i$ favors a 4.3 kms$^{-1}$ solution. We therefore suggest that the low signal-to-noise ratio in the spectra of LEP 1718-4131 is resulting in inflated $v$sin$i$ estimates for some epochs. While the explanation for \hbox{GJ 334 B} is less clear, this star is also faint, with a low cross-correlation coefficient of roughly $h=0.4$ in our two TRES spectra; our $v$sin$i$ measurements may therefore be similarly overestimated. Alternatively, the 2MASS $K$-band magnitude could be inaccurate due to the bright K-dwarf primary at 8" separation biasing the background estimation; indeed, such a possibility is noted for GJ 334 B in the 2MASS catalog \citep{Cutri2003}.

\begin{figure*}[t]
    \centering
    \makebox[\textwidth][c]{\includegraphics[width=\textwidth]{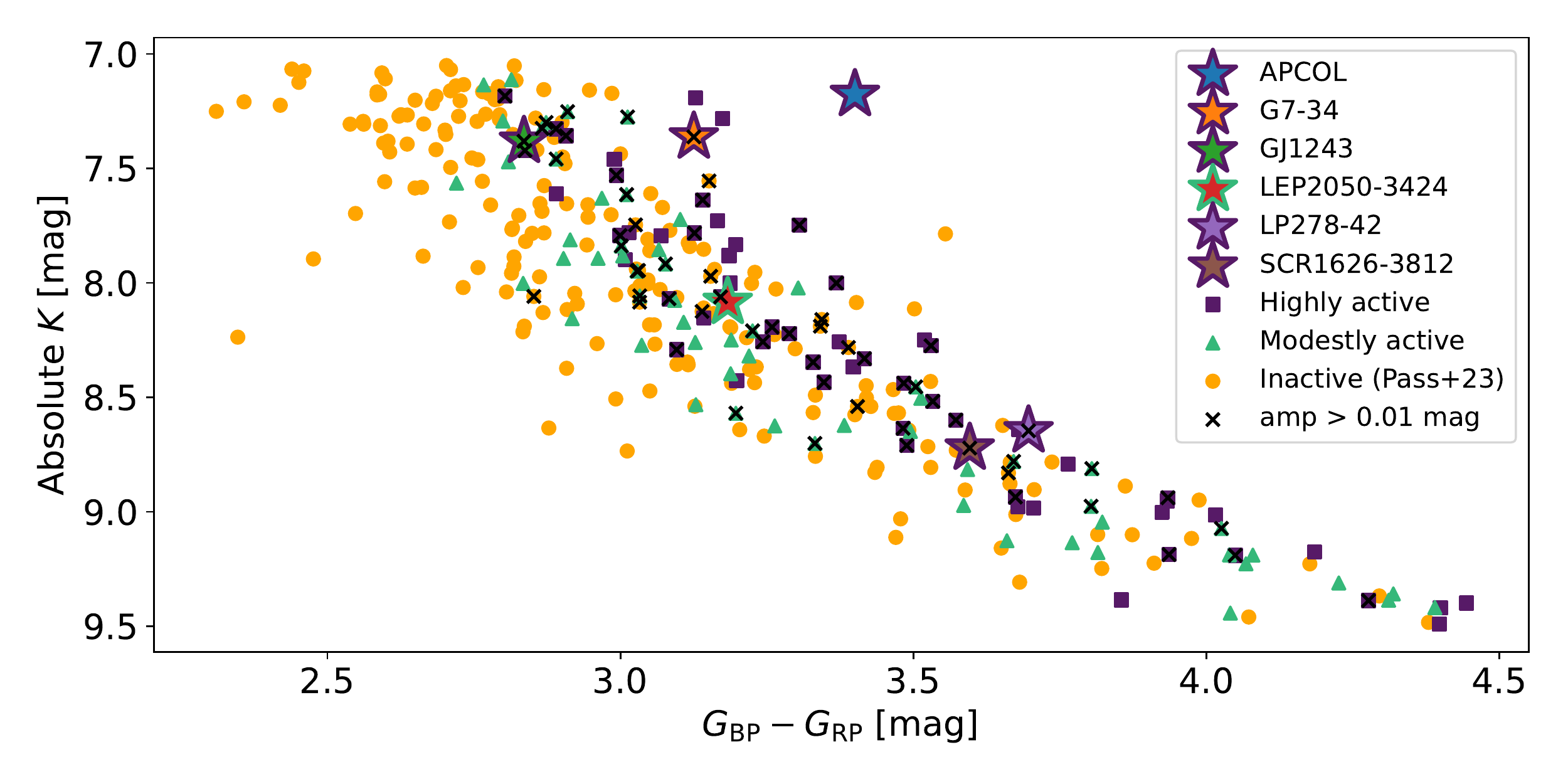}}
    \caption{A color-magnitude diagram of the 15pc sample of 0.1--0.3M$_\odot$ M dwarfs without close binary companions, with magnitudes from Gaia and 2MASS. The 123 active stars are from this work while the 200 inactive stars are described in \citet{Pass2023}. Black xs denote stars with measured peak-to-peak photometric amplitudes exceeding 0.01 mag, including 12 inactive stars with measurements tabulated in \citet{Medina2022}. Candidate members of young moving groups are labeled. At a given $K$-band magnitude, more active stars generally exhibit redder colors. This behavior does not necessarily indicate these stars are overluminous due to youth: within a coeval cluster, more rapidly rotating M dwarfs have been found to be redder, perhaps due to decreased temperatures due to high starspot coverage \citep[e.g.,][]{Covey2016}.}
    \label{fig:cmd}
\end{figure*}

While we have neglected stars with $v < 3.4$ kms$^{-1}$, as they should all have undetectable $v$sin$i$ at the resolution of our spectrographs and therefore be uninformative for this analysis, there is one star for which this is not the case. We measure a median $v$sin$i$ of \hbox{4.1 kms$^{-1}$} for \hbox{WT 84,} yielding an unphysical sin$i$ of 2.47 when compared to its 5.23-day rotation period \citep{Newton2018}. This period is apparent in both MEarth and TESS photometry. The non-zero $v$sin$i$ is detected in all four of our CHIRON spectra, with our measurements ranging from 3.6 to 4.5 kms$^{-1}$. While we have used 3.4 kms$^{-1}$ as our threshold for $v$sin$i$ significance for the purposes of consistency across the sample, this threshold is half a resolution element for the lower-resolution TRES spectrograph. As CHIRON has a resolution of $R=80000$, it is nominally sensitive to $v$sin$i$ down to a lower cutoff of 1.9 kms$^{-1}$; that is to say, a measurement of 4.1 kms$^{-1}$ with CHIRON is well above the threshold for significance. An unresolved binary could possibly lead to an inflated estimate of $v$sin$i$, but we do not identify any statistically significant RV variation between our observations. WT 84 also does not have a high Gaia RUWE that might suggest an astrometric perturbation by a companion. If our measurements of $v$sin$i$ and $P_{\rm rot}$ are accurate, the remaining possibility is an issue with $R_*$. We would require our estimated radius to more than double to resolve the tension between the velocity estimates. As we have discussed above, the \citealt{Benedict2016}+\citealt{Boyajian2012} method employed here should provide a reasonable radius estimate for the stars in our sample. However, this calculation assumes the star is on the main sequence; if WT 84 is a pre-main-sequence star, its radius may be larger than we have assumed, bringing the photometric and spectroscopic velocity estimates into better agreement. We note that WT 84 is very active in H\textalpha, with an even higher level of emission than AP Col, a pre-main-sequence star and member of a young moving group \citep{Riedel2011}. However, WT 84 does not appear elevated on a color-magnitude diagram, which would be expected for a pre-main-sequence star. In addition, the UVW space motion analysis of this star performed in \citet{Medina2022} indicates that while WT 84 is likely a member of the thin disk, it has a large total space velocity compared to the known members of young moving groups studied in that work (48 kms$^{-1}$, as compared to 11, 12, and 15 kms$^{-1}$). We therefore do not have evidence that WT 84 is an exceptionally young star, leaving the puzzle of its rotation unresolved.

As mentioned in the above discussion, our calculation of stellar radius assumes the star is on the main sequence. While we do not have evidence that WT 84 is a pre-main-sequence star, it is worthwhile to consider whether there are potentially other pre-main-sequence stars in the sample for which our radius and inclination estimates would not be valid. We use our radial velocity measurements, Gaia astrometry, and the BANYAN $\Sigma$ tool \citep{Gagne2018} to search for stars in our sample that may be members of young moving groups. This analysis yields six candidate moving-group members: AP Col and GJ 1243, candidate members of the Argus association (40-50Myr; \citealt{Zuckerman2019}) with 99.2\% and 80.2\% probability, respectively; G 7-34, candidate member of the AB Doradus moving group (150Myr) with 99.9\% probability; and LEP 2050-3424, LP 278-42, and SCR 1626-3812, candidate members of the Carina-Near moving group (200Myr) with 78.8\%, 93.8\%, and 96.0\% probability, respectively. In Figure~\ref{fig:cmd}, we note the location of these stars in a color-magnitude diagram of the 15pc sample. Only AP Col is significantly overluminous relative to inactive stars in the sample; as mentioned above, past work has argued that that this star is definitively a member of Argus \citep{Riedel2011}. Our estimated radius and inclination of AP Col are therefore likely to be inaccurate; however, excluding this star from our population-level analyses (such as Figures~\ref{fig:hist},~\ref{fig:amp},~\ref{fig:inclinations}, and~\ref{fig:iso}) does not change any of the conclusions we have presented. G 7-34 may also be slightly overluminous (and was noted as an AB Dor member in \citealt{Bell2015}). This star also has similar magnitudes and colors as GJ 669 B and GJ 896 B, which were not flagged as members of young moving groups but could potentially be young. Our conclusions are robust to the inclusion or exclusion of these three stars. As noted in \citet{Winters2021}, the inactive star GJ 1230 B also appears overluminous on a color-magnitude diagram; it is possible that stars like GJ 669 B, GJ 896 B, and GJ 1230 B with nearby bright primaries have poorly estimated $K$-band magnitudes due to contamination biasing the background estimation, as we discussed as a possibility for GJ 334 B.

Figure~\ref{fig:cmd} also shows that at a given magnitude, active stars generally have redder colors. This result does not necessarily imply that active stars are generally overluminous due to youth: studies of young clusters like the Pleiades have shown that at fixed age, more rapidly rotating M dwarfs have redder colors \citep{Stauffer2003, Kamai2014, Covey2016}, potentially as a result of higher starspot filling fractions on rapidly rotating stars leading to cooler temperatures. Moreover, overluminosity of pre-main-sequence stars cannot reasonably explain Figure~\ref{fig:cmd}. Recall that 123/323=38\% of M dwarfs in our 15pc sample are active. If one assumes that star formation has been constant over the past 8 Gyr and that these stars would be overluminous for 300 Myr \citep[see discussion in][]{Pass2022}, we would expect less than 4\% of our stars to be overluminous. The bias of active stars towards redder colors is a much larger effect than would be expected from contamination of the sample by pre-main-sequence stars.

\section{Radial-velocity variability}
\label{sec:rv}

Our individual RV measurements are given in Table~\ref{tab:rvs}. Ten of our active stars exhibit variation in excess of their nominal uncertainties based on a chi-squared analysis, with a less than 1\% chance that the data are consistent with an unvarying model, i.e., $P(\chi^2) < 1\%$. Given our sample size, one such outlier would be expected due to random chance, on average, in the absence of any additional activity-induced jitter. One of the flagged stars is LHS 252, whose RV variability is likely the result of its two giant planets detected in \citet{Morales2019}, although activity-induced variability may also be contributing to the significance of the signal.

\begin{deluxetable}{lccl}[t]
\tabletypesize{\scriptsize}
\tablecaption{Epoch observations of radial velocity and H\textalpha \label{tab:rvs}}
\tablehead{ 
\colhead{Column} & 
\colhead{Format} &  
\colhead{Units} & 
\colhead{Description}} 
\startdata 
1 & A13 & --- & Star name \\
2 & F4.4 & days & BJD - 2457640 \\
3 & F5.3 & kms$^{-1}$ & Radial velocity \\ 
4 & F4.3 & kms$^{-1}$ & Uncertainty in radial velocity \\ 
5 & F3.1 & \AA\ & Equivalent width of H\textalpha
\enddata
\tablecomments{Full table available in machine-readable form. The uncertainties are internal errors that are appropriate when considering relative radial velocities. If using these data as absolute radial velocities, add an additional 0.5kms$^{-1}$ error in quadrature to account for the uncertainty in the absolute radial velocity of the template. \newline}
\end{deluxetable}

\vspace{-0.45cm}
Aside from LHS 252, three out of the four most variable stars are known activity-induced RV variables; these each have $P(\chi^2)$$\leq$$0.0001$\%. We observe variability of LP 71-82, consistent with the starspot-induced RV variation identified in \citet{Robertson2020}, as well as variability of GJ 51 and G 99-49, consistent with their identification as active, RV-loud stars in \citet{Tal-Or2018}. As activity-induced RV variability varies with both wavelength and time, we do not expect to measure the same amplitude as these previous works; however, we find that our measurements are consistent with the literature within factors of a few.

\begin{deluxetable}{lrrrrrr}[t]
\tabletypesize{\footnotesize}
\tablecolumns{7}
\tablewidth{0pt}
 \tablecaption{Properties of active, RV-loud stars \label{tab:var}}
 \tablehead{
 \colhead{ \vspace{-0.1cm}Name} & 
 \colhead{ \vspace{-0.1cm} $N_{\rm obs}$} &
 \colhead{$v\sin i$ } &
 \colhead{$P_{\rm rot}$} &  
 \colhead{Amp.} & 
 \colhead{Obs.\ var.} &
 \colhead{Pred.\ var.} \\
 \colhead{} &
 \colhead{} &
 \colhead{[kms$^{-1}$]} &
 \colhead{[d]} &
 \colhead{[mag]} &
 \colhead{[kms$^{-1}$]} &
 \colhead{[kms$^{-1}$]}}
\startdata
G 99-49 & 11 & 5.3 & 1.81 & 0.015 & 0.05 & 0.08 \\
GJ 51 & 11 & 10.6 & 1.02 & 0.047 & 0.34 & 0.50 \\
GJ 669 B & 9 & 7.0 & 1.46 & 0.007 & 0.10 & 0.05 \\
LHS 1376 & 4 & 3.9 & 3.02 & 0.014 & 0.07 & 0.05 \\
LHS 1638 & 5 & 6.6 & 1.59 & 0.008 & 0.08 & 0.05 \\
LHS 2320 & 16 & 12.3 & 0.69 & 0.032 & 0.33 & 0.39 \\
LP 71-82 & 12 & 10.2 & 0.28 & 0.010 & 0.15 & 0.11
\enddata
\tablecomments{Amp.\ is the peak-to-peak photometric amplitude measured by MEarth \citep{Newton2016}. Obs.\ var.\ is the sample standard deviation of our $N_{\rm obs}$ observations. Pred.\ var.\ is our order-of-magnitude prediction of the variability, $v$sin$i \times $Amp. \newline}
\end{deluxetable}
\vspace{-0.4cm}

\begin{figure}[h]
    \centering
    \includegraphics[width=\columnwidth]{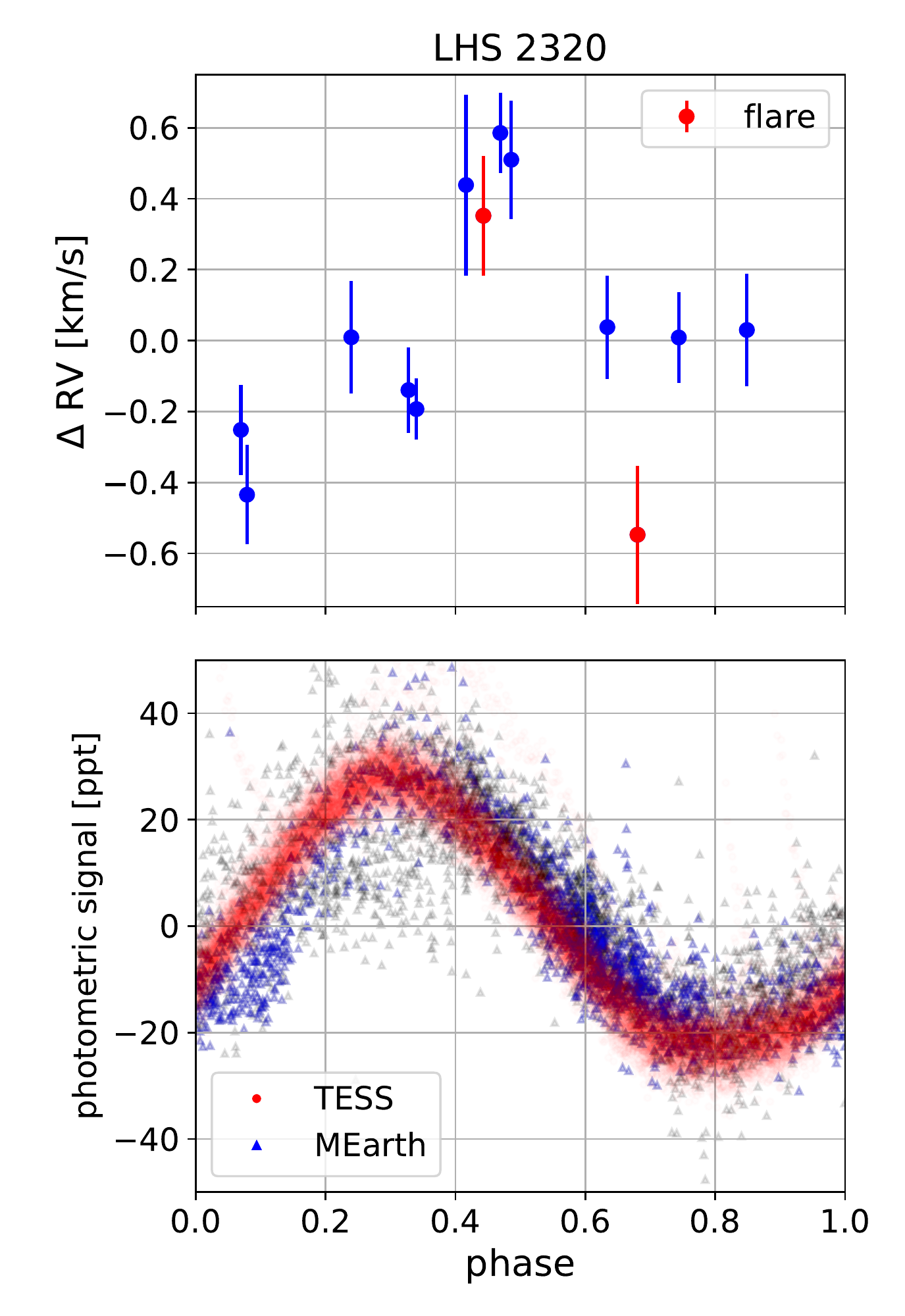}
    \caption{In the upper panel, we show 13 radial velocities of LHS 2320 collected between 2020 Dec and 2021 May, phased to the star's 0.692-day rotation period. Most observations of LHS 2320 have H\textalpha\ equivalent widths between \hbox{-6 and -10\AA}; however, two observations feature H\textalpha\ spikes to nearly -20\AA. These two epochs are indicated in red, as the flares may affect our RV measurements (as in, for example, \citealt{Robertson2020}). Observations appear clustered when phased by the rotation period. In particular, the four highest RV measurements all occur at the same phase. In the lower panel, we show the photometric variability of the star as observed by TESS and MEarth. While MEarth observed this star for over a decade (pale triangles), we highlight the observations that are contemporaneous with the RV observations (blue triangles). The TESS observations offer higher precision but were taken in 2021 Nov/Dec (i.e., they are not contemporaneous). We see that the time of maximal flux corresponds to $\Delta$RV near 0, as expected \citep[e.g., see Figure 11 of][]{Boisse2012}.}
    \label{fig:lhs2320}
\end{figure}

The other highly variable star is LHS 2320, whose RV variability has not been previously studied. We find that LHS 2320 is very RV loud, with a standard deviation of 330 ms$^{-1}$ determined from 16 observations. This is broadly consistent with expectations for activity-induced RV variability based on the photometric amplitude observed with the MEarth array \citep{Newton2016} in a similar optical bandpass to our TRES observations \citep{Berta2012}. To obtain an order-of-magnitude estimate of the expected RV variability due to starspots, we multiply the $v$sin$i$ from TRES with the photometric amplitude from MEarth to find \hbox{12.3 kms$^{-1}$ $\times$ 0.032 = 390 ms$^{-1}$}. We make a similar estimate for LP 71-82, GJ 51, and G 99-49 (Table~\ref{tab:var}), finding that in each case our simplistic estimate of the expected variability is consistent with the observed variability within a factor of two. Furthermore, thirteen of our LHS 2320 observations were taken over a single observing season; we find that the variability over this observing season is in phase with the 0.692-day rotation period seen in MEarth and TESS (Figure~\ref{fig:lhs2320}).

The remaining five candidate variables have 0.0001\% $<$ $P(\chi^2)$ $<$ 1\%. Of these, three stars have measured rotational broadening ($v$sin$i$ $>$ 3.4kms$^{-1}$): \hbox{LHS 1376,} GJ 669 B, and LHS 1638. Again, we find that the observed variability and our simplistic estimate of starspot-induced variation is consistent within a factor of two (Table~\ref{tab:var}). GJ 1224 and LP 731-76 also exhibit variation in excess of nominal uncertainties, although they do not show measurable rotational broadening (and in the case of LP 731-76, lack a rotation period measurement). As we do not have a robust estimate of $v$sin$i$ given the resolution of our spectrographs, the order-of-magnitude estimate described above is not appropriate for these stars. Of course, our observed variation may still be (and is likely to be) the result of activity. True spot patterns are complex and evolving, with flares and chromospheric activity also capable of generating RV signatures \citep[e.g.,][]{Robertson2020}; in this more subtle regime, time-resolved spectroscopic activity indicators are necessary for discriminating between planetary signals and activity-induced variability \citep[e.g.,][]{Lafarga2021}, although our sparse observing strategy does not allow for this type of analysis. Barnard's Star is a particularly illustrative example of the insidiousness of activity: \citet{Lubin2021} found that its planet reported in the literature was an artifact of activity-induced variability, despite this star being a slow rotator and inactive in H\textalpha.

\begin{figure}[h]
    \centering
    \includegraphics[width=\columnwidth]{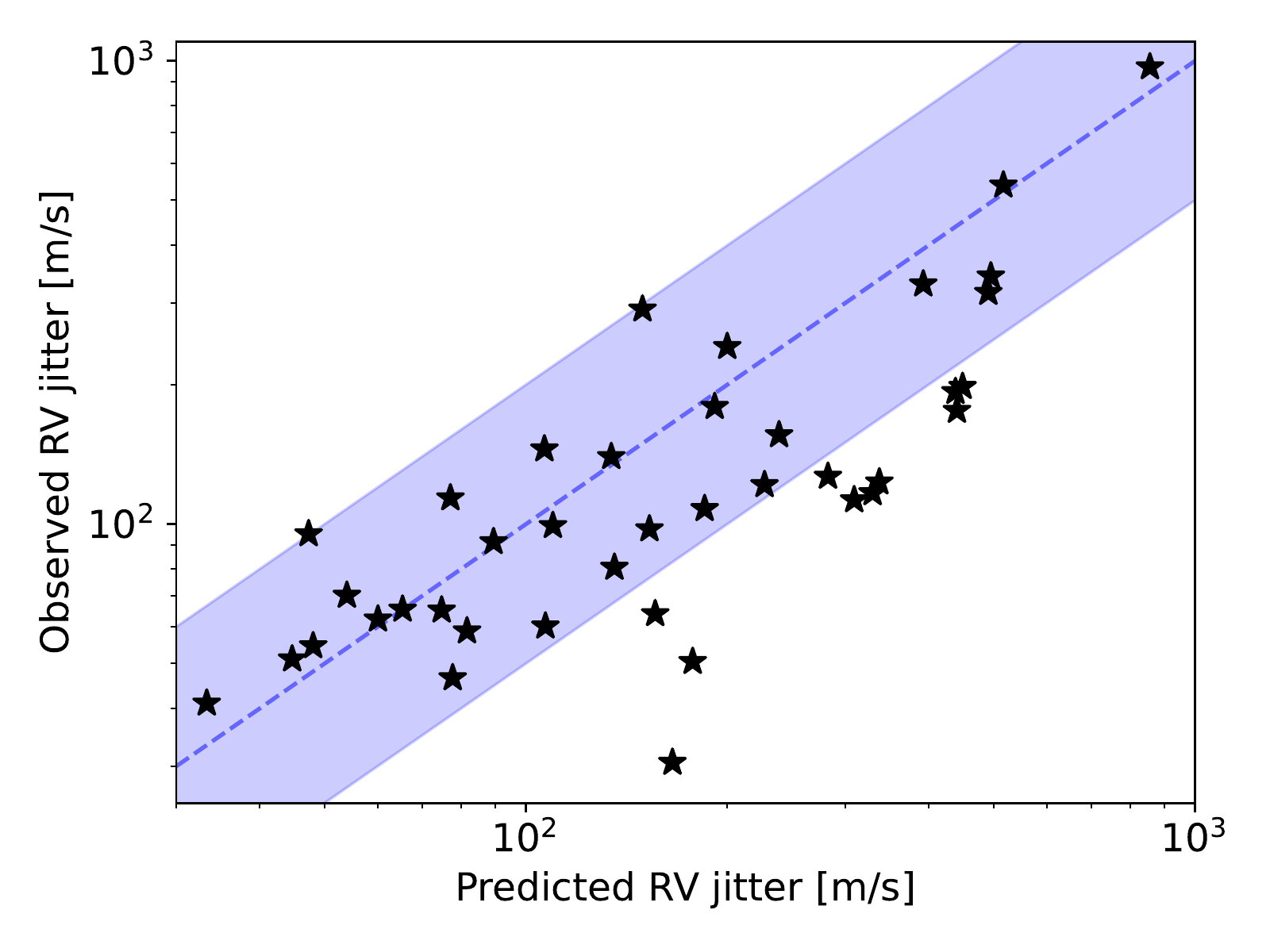}
    \caption{We plot our simple prediction of RV jitter: the photometric amplitude multiplied by $v$sin$i$. This estimate correlates with the sample standard deviation we observe in our RV time series. The blue line indicates unity, with the shaded region showing agreement within a factor of 2. We show the 39 stars with resolved rotational broadening ($v$sin$i >$ 3.4kms$^{-1}$), measured photometric amplitudes, and whose predicted RV jitter is larger than that star's median RV uncertainty.}
    \label{fig:ampl}
\end{figure}

There are stars with large $v$sin$i$ and large photometric amplitudes that are not variable at the $P(\chi^2) < 1\%$ level. There are many reasons why this may be the case: with only four spectroscopic observations, we may be sampling similar phases of the rotation period by chance; spot patterns change over time, and so the literature photometric amplitude may not correspond to the amplitude at the time of the RV observations; the presence of bright faculae as opposed to dark spots may lead to a different velocity structure of the features; the observational uncertainties may be large relative to the predicted RV amplitude. This latter case is significant, as RV uncertainties are inversely proportional to the information content in the spectrum, $Q$; therefore, stars with larger $v$sin$i$ will experience greater rotational broadening, lower $Q$, and larger uncertainties \citep{Bouchy2001}. This appears to be the case for APM 0237-5928, which has a MEarth amplitude of 0.025 mag \citep{Newton2018} and a $v$sin$i$ of 20.5 kms$^{-1}$. Our simple relation predicts activity-induced variability of 510 ms$^{-1}$ and we observe a standard deviation of 540 ms$^{-1}$, in line with this estimate. However, the large $v$sin$i$ results in large uncertainties on the individual radial velocities, preventing the signal from meeting our significance threshold. In this case, we measure $P(\chi^2) = 1.7$\%. Another example is GJ 1167, with a $v$sin$i$ of 45.5 kms$^{-1}$, a MEarth amplitude of 0.019 \citep{Newton2016}, a predicted variability of 860 ms$^{-1}$, a standard deviation of 960 ms$^{-1}$, and $P(\chi^2) = 1.1$\%. While we do not always flag stars with large $v$sin$i$ and large photometric amplitude as statistically significant variables, we nonetheless find that RV variability generally correlates with our simple prediction in the sample at large (Figure~\ref{fig:ampl}).

\section{Summary}
\label{sec:conclusion}
We gathered multi-epoch, high-resolution spectroscopic observations for all 0.1-0.3M$_\odot$ M dwarfs within 15pc, a population described in \citet{Winters2021}. After omitting close binaries, this sample consists of 323 stars. From these spectra, we measured H\textalpha\ emission using the method of \citet{Medina2020} and defined the active subsample as the 123 stars with median H\textalpha\ emission stronger than -1\AA; this active subsample represents 38\% of the stars without close binary companions. We report properties of these stars in Table~\ref{tab:summary}, including H\textalpha\ equivalent width, $v$sin$i$, rotation period, inclination, and the significance of any RV variation.

We report rotation periods for all but eight stars in the sample, including 31 new detections using TESS and/or MEarth. By combining these rotation periods with our $v$sin$i$ measurements, we measure inclinations for the 90 stars in the sample with $v$ $>$ 3.4 kms$^{-1}$ (and therefore, with measurable $v$sin$i$ given the resolution of our spectrographs). The distribution of inclinations is relatively consistent with expectations for an isotropic distribution of spin axes, but with a lack of pole-on stars. We hypothesize that pole-on stars are among those with missing rotation periods, resulting from the decreased photometric amplitude expected in this geometry.

Our sample is volume complete, allowing us to draw conclusions about the overall distribution of M dwarfs. We find that 92$\pm$3\% of active, low-mass M dwarfs exist in the rapidly rotating mode with $P_{\rm rot} < 20$ days, with the majority (74$\pm$4\%) having rotation periods less than 2 days. Among the 8$\pm$3\% with rotation periods longer than 20 days, we identify two subpopulations: there are low-mass M dwarfs with periods of 20-40 days with high levels of H\textalpha\ emission, which we hypothesize are currently undergoing rapid spindown and transitioning between the rapidly and slowly rotating modes. There are also active stars with rotation periods at the short end of the slowly rotating sequence and with modest H\textalpha\ emission, which we interpret as M dwarfs that have newly spun down to the slowly rotating mode.

We observe a correlation between H\textalpha\ emission and rotation within the rapidly rotating mode, with $P_{\rm rot} < 0.5$-day rotators typically having greater H\textalpha\ emission than stars with $2 < P_{\rm rot} < 10$ days. We also find that the M dwarfs with the greatest H\textalpha\ emission tend to have the largest photometric amplitudes; this correlation does not persist for modestly active M dwarfs, whose distribution of amplitudes is statistically equivalent to the distribution for inactive M dwarfs. Our observed correlation between rotation and activity may actually be a correlation between activity and age; in \citet{Pass2022}, we showed that low-mass M dwarfs slowly spin down within the rapidly rotating mode, reaching periods of 2--10 days by ages of a few gigayears before rapidly spinning down to the slowly rotating mode at more advanced ages. That said, there is a large dispersion in the rotation periods of pre-main-sequence stars in the post-disk-locking phase: a low-mass M dwarf with a period of 5 days could be a very young star with slow initial rotation, or it could be a few gigayears old and slowly spinning down from a faster initial rate. This inherent dispersion will complicate attempts to use fully convective M dwarfs for gyrochronology. We also find that the rotation periods of field M dwarfs in the rapidly rotating mode have a mass-dependent slope, with 0.1M$_\odot$ stars rotating more rapidly on average than 0.3M$_\odot$ stars; previous studies of young clusters indicate that this trend is imprinted on the population in the pre-main-sequence phase \citep{Somers2017}. 

Lastly, we report our multi-epoch RV and H\textalpha\ measurements and discuss activity-induced RV variability. Seven stars in our sample show highly significant variability that we ascribe to spot-induced variation using a simple model, where we estimate RV variation as the product of $v$sin$i$ and the amplitude of photometric modulation. In each case, this simple prediction is consistent with the observed variability within a factor of two. We also find that this simple estimate can explain the RV jitter we observe in the sample at large. This excess noise intrinsic to active M dwarfs motivates our exclusion of the active sample from our search for planets in \citet{Pass2023}.

\section*{Acknowledgements}
We thank the TRES and CHIRON teams for their support, including Perry Berlind, Allyson Bieryla, Lars Buchhave, Michael Calkins, Gilbert Esquerdo, Pascal Fortin, Todd Henry, Hodari James, Wei-Chun Jao, David Latham, Jessica Mink, Leonardo Paredes, Samuel Quinn, Andrew Szentgyorgyi, and Andrei Tokovinin. We also thank Evgenya Shkolnik for helpful discussions on M-dwarf activity and the anonymous referee for comments that improved this manuscript.

EP is supported in part by a Natural Sciences and Engineering Research Council of Canada (NSERC) Postgraduate Scholarship. This work is made possible by a grant from the John Templeton Foundation. The opinions expressed in this publication are those of the authors and do not necessarily reflect the views of the John Templeton Foundation.  This material is based upon work supported by the National Aeronautics and Space Administration under grants 80NSSC19K0635, 80NSSC19K1726, 80NSSC21K0367, 80NSSC22K0165, and 80NSSC22K0296 in support of the TESS Guest Investigator Program and grant 80NSSC18K0476 issued through the XRP program.

This paper includes data collected by the TESS mission, which are publicly available from the Mikulski Archive for Space Telescopes (MAST). Funding for the TESS mission is provided by the NASA's Science Mission Directorate. The MEarth Project acknowledges funding from the David and Lucile Packard Fellowship for Science and Engineering, and the National Science Foundation under grants AST-0807690, AST-1109468, AST-1616624 and AST-1004488 (Alan T. Waterman Award). This work has made use of data from the European Space Agency (ESA) mission
{\it Gaia} (\url{https://www.cosmos.esa.int/gaia}), processed by the {\it Gaia}
Data Processing and Analysis Consortium (DPAC,
\url{https://www.cosmos.esa.int/web/gaia/dpac/consortium}). Funding for the DPAC
has been provided by national institutions, in particular the institutions
participating in the {\it Gaia} Multilateral Agreement. This publication also makes use of data products from the Two Micron All Sky Survey, which is a joint project of the University of Massachusetts and the Infrared Processing and Analysis Center/California Institute of Technology, funded by the National Aeronautics and Space Administration and the National Science Foundation.

%

\facilities{CTIO:1.5m (CHIRON), FLWO:1.5m (TRES), MEarth, TESS}
\software{\texttt{Astropy} \citep{AstropyCollaboration2013, AstropyCollaboration2018}, \texttt{Lightkurve} \citep{LightkurveCollaboration2018}, \texttt{Matplotlib} \citep{Hunter2007}, \texttt{NumPy} \citep{Harris2020}, \texttt{pandas} \citep{Reback2021}, \texttt{PyMC3} \citep{Salvatier2016}, \texttt{SciPy} \citep{Virtanen2020}}

\bibliography{active}{}
\bibliographystyle{aa_url}



\end{document}